\documentclass[aps,preprint,showpacs,superscriptaddress,prc,floatfix,amssymb,amsfonts]{revtex4}%

\usepackage{graphicx}
\usepackage{amsmath}
\usepackage{amsfonts}
\usepackage{amssymb}
\usepackage{longtable}

\begin{document}

\preprint{APS/123-QED}

\title{High-spin intruder band in $^{107}$In}%

\author{E.~Ideguchi}
\affiliation{%
Center for Nuclear Study, the University of Tokyo, 
Wako, Saitama 351-0198, Japan
}
\email{ideguchi@cns.s.u-tokyo.ac.jp}

\author{B.~Cederwall}%
\author{E.~Ganio{\u g}lu}
 \altaffiliation[Also at ]{Department of Physics, Faculty of Science, 
Istanbul University}%
\author{B.~Hadinia}
\author{K.~Lagergren}
\author{T.~B\"{a}ck}
\author{A.~Johnson}
\author{R.~Wyss}
\affiliation{%
Department of Physics, The Royal Institute of Technology, 
SE-10691 Stockholm, Sweden
}%

\author{S.~Eeckhaudt}
\author{T.~Grahn}
\author{P.~Greenlees}
\author{R.~Julin}
\author{S.~Juutinen}
\author{H.~Kettunen}
\author{M.~Leino}
\author{A.-P.~Leppanen}
\author{P.~Nieminen}
\author{M.~Nyman}
\author{J.~Pakarinen}
\author{P.~Rahkila}
\author{C.~Scholey}
\author{J.~Uusitalo}
\affiliation{
Department of Physics, University of Jyv\"askyl\"a, P.O. Box 35,
FI-40014 Jyv\"askyl\"a, Finland
}%

\author{D.~T.~Joss}
\affiliation{%
CCLRC, Daresbury Laboratory, Daresbury, Warrington WA4 4AD, United Kingdom
}%

\author{E.~S.~Paul}
\author{D.~R.~Wiseman}
\affiliation{%
Oliver Lodge Laboratory, Department of Physics, University of Liverpool,
Liverpool L69 7ZE, United Kingdom
}%

\author{R.~Wadsworth}
\affiliation{%
Department of Physics, University of York, 
Heslington, York YO10 5DD, United Kingdom
}%

\author{A.\ V.\ Afanasjev}
\affiliation{Department of Physics and Astronomy, Mississippi State 
University, Mississippi State, Mississippi 39762, USA}

\author{I.\ Ragnarsson}
\affiliation{Division of Mathematical Physics, LTH, Lund University, 
P.O. Box 118 SE-221 00 Lund, Sweden}

\date{\today}

\begin{abstract}
  High-spin states in the neutron deficient nucleus $^{107}$In were 
studied via the $^{58}$Ni($^{52}$Cr, 3p) reaction. 
In-beam $\gamma$ rays were measured using the JUROGAM detector array. 
A rotational cascade consisting of ten $\gamma$-ray transitions 
which decays to the 19/2$^{+}$ level at 2.002~MeV was observed. 
The band exhibits the features typical for smooth terminating bands 
which also appear in rotational bands of heavier nuclei in the 
A$\sim$100 region. 
The results are compared with Total Routhian Surface and Cranked 
Nilsson-Strutinsky calculations.
\end{abstract}

\pacs{
23.20.Lv, 24.60.Dr, 23.20.En, 27.60.+j
}

\maketitle

\section{Introduction}
%
%
The structure of nuclei close to $^{100}$Sn has received increasing 
attention in recent years. 
Excited states of neutron deficient nuclei with Z$\sim$50 are expected 
to be predominantly of single-particle nature at low spin due to the 
presence of a spherical shell gap for protons. 
However, recent experimental and theoretical investigations have elucidated 
additional important excitation mechanisms, such as magnetic rotation 
\cite{magrot}, and deformed rotational bands of dipole and quadrupole 
character exhibiting smooth band  termination  
\cite{smooth1,sb109,Smooth-PR,smooth-dipole,sn108,Te110}. 
The diversity of excitation modes in these nuclei make them particularly 
interesting systems to study.

In the A$\sim$100 region, the observation of well-deformed structures are 
interpreted as being based on 1p-1h \cite{smooth-dipole,Te110} and 2p-2h 
\cite{smooth1,Smooth-PR,sb109,sn108} proton excitations across the Z=50 
closed-shell gap and on the occupation of the intruder $h_{11/2}$ orbital.
These studies have in particular high-lighted the so-called 
``smooth band termination'' phenomenon, following alignment of the valence 
nucleons outside the $^{100}$Sn doubly-magic core.  
In the previous studies, levels in $^{107}$In were extended up to 
$I$= (33/2) at 6.893~MeV, 
the last member of a magnetic-dipole band structure \cite{in107_prc58}.
A well deformed intruder structure has been observed earlier in $^{107}$In 
(see Fig.\ 23 in Ref.\ \cite{Smooth-PR}), 
but its detailed analysis has not been published so far.

In order to further investigate the existence of well-deformed structures in 
the A$\approx$100 mass region, we have obtained new data on the high-spin 
states of $^{107}$In.
In the present work, a rotational band was observed, extending to high 
angular momentum ($> 30 \hbar$).
It is connected with the low-lying yrast states via 
several $\gamma$-ray transitions. 

%
%
A rotational-like level structure in $^{107}$In has recently been reported 
\cite{in107_epja43}.
Although some of the $\gamma$-ray transitions are identical to those observed 
in the present work, the results are generally not in agreement with the results 
obtained in the present work. 
The band was not observed to lower spin states and it was not connected to 
the lower-lying states in $^{107}$In.

\begin{figure*}[ht]
\centering
\includegraphics[width=16cm]{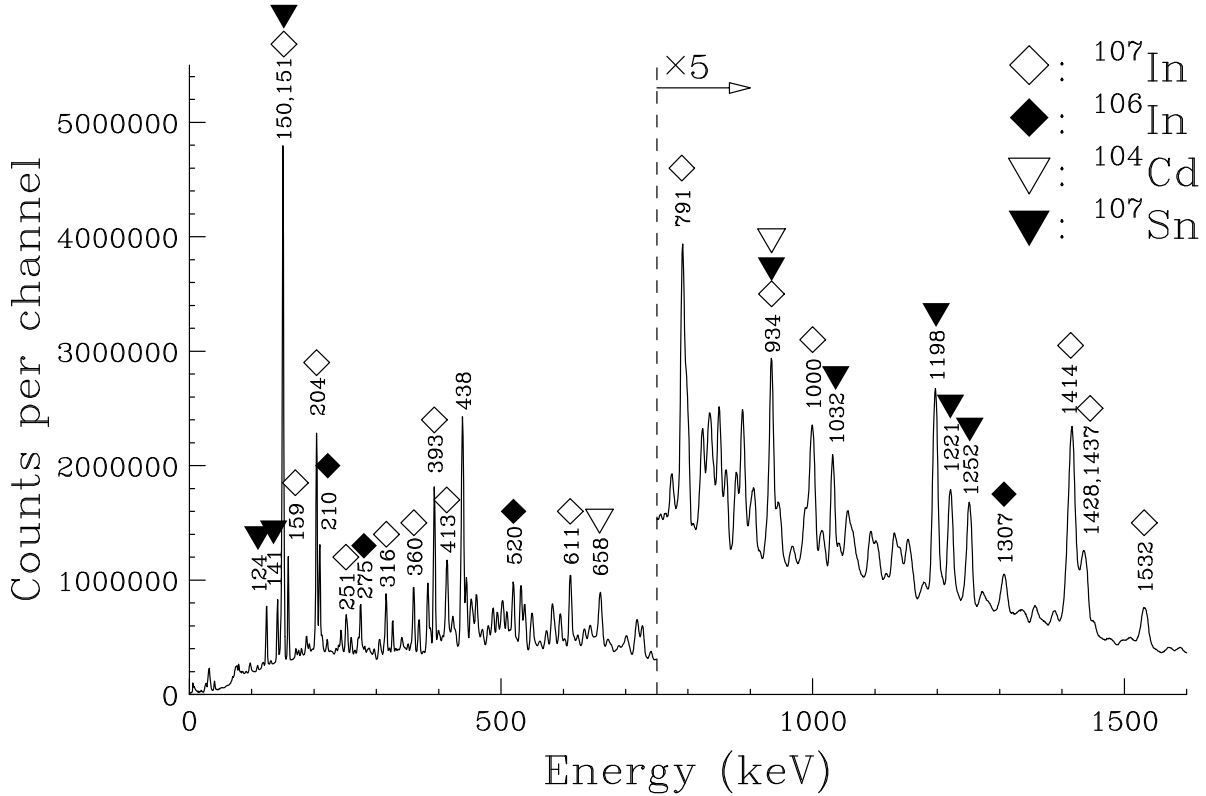}
\caption{
Total $\gamma-\gamma$ coincidence spectrum.
Gamma-ray peak energies are labeled in keV for strong peaks. 
Previously known transitions in $^{107}$In as well as
those from the other principal evaporation residues, 
$^{104}$Cd, $^{106}$In, and $^{107}$Sn are observed as indicated 
in the figure.}
\label{proj}
\end{figure*}

\section{Experimental details}
%
%
The experiment was performed at the JYFL accelerator facility at the
University of Jyv\"askyl\"a, Finland.
%
%
The $^{52}$Cr ions, accelerated by the JYFL $K$ $=$ 130~MeV cyclotron to 
an energy of 187~MeV, were used to bombard a target consisting of two stacked
self-supporting foils of isotopically enriched (99.8\%) $^{58}$Ni.
The targets were of thickness 580~$\mu$g/cm$^2$ and 640~$\mu$g/cm$^2$.
The average beam intensity was 4.4~particle-nA during 5~days of irradiation 
time.
%
%
High-spin states in $^{107}$In were populated by the fusion-evaporation
reaction $^{58}$Ni($^{52}$Cr, 3p)$^{107}$In.
%
%
Prompt $\gamma$-rays were detected at the target position by the JUROGAM
$\gamma$-ray spectrometer consisting of 43~EUROGAM \cite{eurogam}
type escape-suppressed high-purity germanium detectors.
In this configuration, JUROGAM had a total photopeak efficiency of about
4.2\% at 1.3~MeV.
Fig.~\ref{proj} shows a total $\gamma-\gamma$ coincidence spectrum.
Gamma-ray peaks of $^{107}$In are observed as belonging to the strongest 
fusion-evaporation channel (3p). 
Other reaction channels such as $^{104}$Cd (1$\alpha$2p), $^{106}$In (3p1n), 
and $^{107}$Sn (2p1n) are also observed clearly.

%
%
The fusion-evaporation products were separated in flight from the beam
particles using the gas-filled recoil separator RITU
\cite{ritu1,ritu2} and implanted into the two double-sided
silicon strip detectors (DSSSD) of the GREAT \cite{great} spectrometer.
The GREAT spectrometer is a composite detector system containing, in
addition to the DSSSDs, a multiwire proportional counter (MWPC), an array
of 28 Si PIN photodiode detectors, and a segmented planar Ge detector.
Each DSSSD has a total active area of 60$\times$40~mm$^2$ and a strip
pitch of 1~mm in both directions yielding in total 4800~independent
pixels.
In this measurement, GREAT was used to filter the events such that recoils 
were separated and transported to the final focal plane of RITU. 

\begin{figure*}[ht]
\centering
\includegraphics[width=16cm]{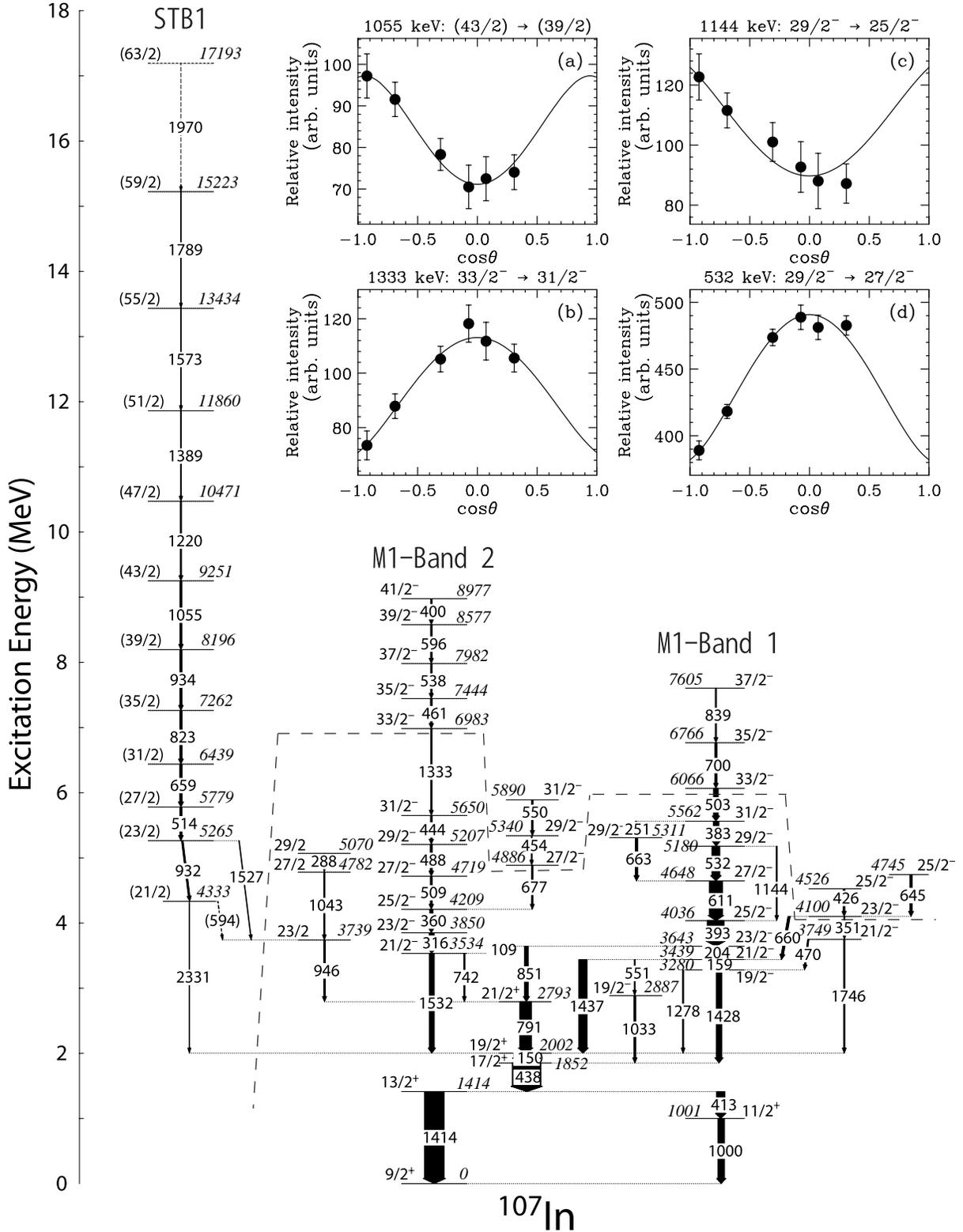}
\caption{A level scheme of $^{107}$In constructed in the present study.
Transitions below dashed lines were previously identified \cite{in107_prc58}.
Insets (a), (b), (c), (d) represent the typical angular distribution of 
$\gamma$ transitions observed with present setup.
Solid lines are fitted curve of the angular distribution function W($\theta$).
}
\label{level_scheme}
\end{figure*}

%
%
The signals from all detectors were recorded independently and provided
with an absolute ``time stamp'' with an accuracy of 10~ns using the
total data readout (TDR) \cite{tdr} acquisition system.
Events associated with the recoil hitting DSSSD of GREAT and 
prompt $\gamma$ rays detected by JUROGAM were sorted offline using GRAIN analysis 
package \cite{GRAIN} to store the recoil-gated multifold $\gamma$ coincidence 
data. 
In total 5.3$\times$10$^{8}$ events were accumulated.
The data were analyzed using the RADWARE data analysis software package 
\cite{radware}.
The multifold event data were sorted offline into an E$_\gamma$-E$_\gamma$
correlation matrix and an E$_\gamma$-E$_\gamma$-E$_\gamma$ cube.
Based on single and double gating on the matrix and cube, respectively,
coincidence relations between observed $\gamma$ rays were examined.
Fig.~\ref{level_scheme} shows a level scheme constructed in the present study.

%
%
In order to determine the $\gamma$-ray multipolarities, 
an angular distribution analysis was performed.
The germanium detectors of the JUROGAM spectrometer were distributed 
over six angles relative to the beam direction with five detectors at 
158$^\circ$, ten at 134$^\circ$, ten at 108$^\circ$, five at 94$^\circ$, 
five at 86$^\circ$, and eight at 72$^\circ$.
Recoil gated coincidence matrices were sorted such that the energies of 
$\gamma$ rays detected at specified angles of JUROGAM, E$_{\gamma}$($\theta$), 
were incremented on one axis, while the energies of coincident $\gamma$ rays 
detected at any angles, E$_{\gamma}$(any), were incremented on the other axis. 
Six angular distribution matrices corresponding to the angles 
$\theta = 158^{\circ}, 134^{\circ}, 94^{\circ}, 86^{\circ}$, and $72^{\circ}$ 
were created. 
Background-subtracted angle-dependent spectra were created by gating on
transitions on the E$_{\gamma}$(any) of the matrices.
Peak areas for a given coincident transition were measured and normalized
by the number of detectors as well as efficiencies at each angle, then fitted 
to the angular distribution function 
W($\theta$) $=$ a$_{0}$(1 + a$_{2}$P$_{2}$(cos$\theta$) 
+ a$_{4}$P$_{4}$(cos$\theta$)).
In the case of stretched quadrupole (E2) transitions, a$_{2}$ and a$_{4}$ 
coefficients will have positive and small negative values, respectively, 
while those of pure stretched dipole (M1 or E1) transitions will be negative 
and zero values, respectively.
%

\section{Results}
%

High-spin levels in $^{107}$In were previously reported up to the ($I$ = 33/2)
state at 6.983~MeV excitation energy \cite{in107_prc58}.
%
%
The spin-parity of the ground state is 9/2$^+$, arising from its $\pi g_{9/2}$
hole character.
%
%
The 19/2$^{+}$ state at 2.002~MeV and the 17/2$^{+}$ state at 1.852~MeV have 
been reported to have isomeric character with half-lives of 0.6(2) and 
1.7(3)~ns, respectively \cite{in107_isomer}.
These lifetimes are long enough so that the $\gamma$-ray emission depopulating 
these states or lower-lying states which are fed via these states, on average 
occurs several cm downstream from the target.
This affects the relative efficiency of detecting such $\gamma$ rays depending on 
the detector angle relative to the beam and hence their angular distribution. 
An attenuation of the alignment could also be affected by the level lifetimes.
Indeed, the angular distributions of transitions below the state at  1.852~MeV
were observed to be isotropic.
However, the angular distribution of the 150~keV transition below the 0.6~ns 
isomer at 2.002~MeV was not isotropic. 
This is consistent with the fact that the half-life of the isomer is shorter 
than that of the 1.852~MeV level.
The half-lives of these two isomers could not be confirmed in this analysis
since the experimental setup was not sensitive for such short lifetimes.

\begin{longtable*}[htbp]{p{2cm}p{2cm}p{2cm}p{2.5cm}p{2.5cm}p{1.5cm}p{1.5cm}p{1.5cm}}
\caption{\label{gamma_table} Gamma rays assigned to $^{107}$In}\\
\hline
\hline
$E_{\gamma}$$^{a}$ & 
\begin{minipage}{1.5cm}
\vspace*{3mm}
$I^{(rel.)}{\gamma}^{b}$ 
\end{minipage}
& $E_i$$^{c}$ & 
\begin{minipage}{1.5cm}
\vspace*{3mm}
\begin{center}
$a_2$ 
\end{center}
\end{minipage}
& 
\begin{minipage}{1cm}
\vspace*{3mm}
\begin{center}
$a_4$  
\end{center}
\end{minipage}
& 
\begin{minipage}{1.5cm}
\vspace*{3mm}
\hspace*{2mm}
$J^{\pi}_i$ 
\end{minipage}
& 
\begin{minipage}{1cm}
\vspace*{3mm}
$\rightarrow$ 
\end{minipage}
& 
\begin{minipage}{1.5cm}
\vspace*{3mm}
\hspace*{2mm}
$J^{\pi}_f$ 
\end{minipage}
\\

   (keV)     &                       & (keV) &       &        &             &               &            \\
\hline
\endhead
\hline
\endfoot
\multicolumn{8}{l}{
\begin{minipage}{10cm}\vspace*{2mm}
{\footnotesize $^{a}$Transition energies accurate to within $\pm0.5\,$keV.}\\
{\footnotesize $^{b}$Intensities are normalized to 1000 for the 438~keV $\gamma$ transition.}\\
{\footnotesize $^{c}$Excitation energies of initial states of $\gamma$ transitions.}\\
\end{minipage}
}
\endlastfoot
  109.3  & 14(2)    &  3643 & $ -0.23(17)  $  & $ 0.24(23)    $& $23/2^{-}$ & $\rightarrow$ & $21/2^{-}$ \\
  149.8  & 768(48)  &  2002 & $ -0.260(15) $  & $ 0.025(22)   $& $19/2^{+}$ & $\rightarrow$ & $17/2^{+}$ \\
  158.6	 & 183(12)  &  3439 & $ -0.266(16) $  & $ 0.003(22)   $& $21/2^{-}$ & $\rightarrow$ & $19/2^{-}$ \\
  204.0	 & 437(26)  &  3643 & $ -0.280(15) $  & $ 0.065(22)   $& $23/2^{-}$ & $\rightarrow$ & $21/2^{-}$ \\
  251.2	 & 50(3)    &  5562 & $ -0.242(47) $  & $ 0.132(69)   $& $31/2^{-}$ & $\rightarrow$ & $29/2^{-}$ \\
  288.4	 & 17(2)    &  5070 & $ -0.36(12)  $  & $ 0.10(17)    $&  29/2      & $\rightarrow$ &  27/2      \\
  315.6	 & 175(12)  &  3850 & $ -0.224(21) $  & $ 0.025(31)   $& $23/2^{-}$ & $\rightarrow$ & $21/2^{-}$ \\
  351.4	 & 33(3)    &  4100 & $ -0.30(18)  $  & $ -0.08(28)   $& $23/2^{-}$ & $\rightarrow$ & $21/2^{-}$ \\
  359.6	 & 115(8)   &  4209 & $ -0.236(27) $  & $ 0.010(40)   $& $25/2^{-}$ & $\rightarrow$ & $23/2^{-}$ \\
  382.9	 & 194(12)  &  5562 & $ -0.272(14) $  & $ 0.070(21)   $& $31/2^{-}$ & $\rightarrow$ & $29/2^{-}$ \\
  393.0	 & 580(36)  &  4036 & $ -0.219(25) $  & $ 0.024(37)   $& $25/2^{-}$ & $\rightarrow$ & $23/2^{-}$ \\
  400.2	 & 53(4)    &  8977 & $ -0.244(31) $  & $ 0.011(50)   $& $41/2^{-}$ & $\rightarrow$ & $39/2^{-}$ \\
  413.2	 & 289(22)  &  1414 & $ -0.101(62) $  & $ 0.117(88)   $& $13/2^{+}$ & $\rightarrow$ & $11/2^{+}$ \\
  426.3	 & 59(6)    &  4526 & $ -0.484(51) $  & $ -0.004(74)  $& $25/2^{-}$ & $\rightarrow$ & $23/2^{-}$ \\
  438.2	 & 1000(60) &  1852 & $  0.071(31) $  & $ 0.068(45)   $& $17/2^{+}$ & $\rightarrow$ & $13/2^{+}$ \\
  443.6	 & 62(5)    &  5654 & $ -0.383(29) $  & $ 0.154(43)   $& $31/2^{-}$ & $\rightarrow$ & $29/2^{-}$ \\
  454.4	 & 33(3)    &  5340 & $ -0.218(40) $  & $ 0.076(59)   $& $29/2^{-}$ & $\rightarrow$ & $27/2^{-}$ \\
  460.5	 & 63(4)    &  7444 & $ -0.206(2)  $  & $ 0.056(2)    $& $35/2^{-}$ & $\rightarrow$ & $33/2^{-}$ \\
  469.6	 & 48(6)    &  3749 & $ -0.094(38) $  & $  0.030(55)  $& $21/2^{-}$ & $\rightarrow$ & $19/2^{-}$ \\
  488.2	 & 75(6)    &  5207 & $ -0.204(38) $  & $ -0.025(56)  $& $29/2^{-}$ & $\rightarrow$ & $27/2^{-}$ \\
  503.4	 & 167(10)  &  6066 & $ -0.237(52) $  & $ -0.053(78)  $& $33/2^{-}$ & $\rightarrow$ & $31/2^{-}$ \\
  509.3	 & 69(6)    &  4719 & $ -0.238(41) $  & $  0.140(60)  $& $27/2^{-}$ & $\rightarrow$ & $25/2^{-}$ \\
  514.3	 & 75(8)    &  5779 & $  0.276(59) $  & $ -0.109(87)  $& (27/2)     & $\rightarrow$ & (23/2)     \\
  532.1	 & 276(18)  &  5180 & $ -0.180(1)  $  & $  0.039(2)   $& $29/2^{-}$ & $\rightarrow$ & $27/2^{-}$ \\
  537.7	 & 54(4)    &  7982 & $ -0.286(34) $  & $  0.188(50)  $& $37/2^{-}$ & $\rightarrow$ & $35/2^{-}$ \\
  550.1	 & 58(5)    &  5890 & $ -0.242(33) $  & $  0.138(48)  $& $31/2^{-}$ & $\rightarrow$ & $29/2^{-}$ \\
  551.1	 & 31(4)    &  3439 & $            $  & $             $& $21/2^{-}$ & $\rightarrow$ & $19/2^{-}$ \\
  593.8  & 5(3)     &  4333 &                 &                &  21/2      & $\rightarrow$ &  23/2      \\
  595.5	 & 56(4)    &  8577 & $ -0.325(14) $  & $  0.025(21)  $& $39/2^{-}$ & $\rightarrow$ & $37/2^{-}$ \\
  611.3	 & 463(28)  &  4648 & $ -0.131(31) $  & $ -0.017(45)  $& $27/2^{-}$ & $\rightarrow$ & $25/2^{-}$ \\
  644.5	 & 87(8)    &  4745 & $ -0.187(24) $  & $ -0.063(35)  $& $25/2^{-}$ & $\rightarrow$ & $23/2^{-}$ \\
  659.0	 & 90(8)    &  6439 & $  0.278(36) $  & $ -0.126(53)  $& (31/2)     & $\rightarrow$ & (27/2)     \\
  659.7	 & 83(10)   &  4100 & $            $  & $             $& $23/2^{-}$ & $\rightarrow$ & $21/2^{-}$ \\
  663.2	 & 89(8)    &  5311 & $ -0.225(35) $  & $  0.048(55)  $& $29/2^{-}$ & $\rightarrow$ & $27/2^{-}$ \\
  676.6	 & 52(5)    &  4886 & $ -0.03(10)  $  & $  0.11(16)   $& $27/2^{-}$ & $\rightarrow$ & $25/2^{-}$ \\
  700.4	 & 73(6)    &  6766 & $ -0.291(60) $  & $ -0.067(93)  $& $35/2^{-}$ & $\rightarrow$ & $33/2^{-}$ \\
  741.5	 & 45(4)    &  3534 & $  0.243(61) $  & $ -0.155(93)  $& $21/2^{-}$ & $\rightarrow$ & $21/2^{+}$ \\
  791.0	 & 419(28)  &  2793 & $ -0.122(31) $  & $  0.023(52)  $& $21/2^{+}$ & $\rightarrow$ & $19/2^{+}$ \\
  823.1	 & 81(10)   &  7262 & $  0.242(17) $  & $ -0.124(68)  $& (35/2)     & $\rightarrow$ & (31/2)     \\
  838.8	 & 45(4)    &  7605 & $ -0.179(34) $  & $  0.040(52)  $& $37/2^{-}$ & $\rightarrow$ & $35/2^{-}$ \\
  850.6	 & 133(10)  &  3643 & $ -0.296(4)  $  & $  0.038(6)   $& $23/2^{-}$ & $\rightarrow$ & $21/2^{+}$ \\
  932.4	 & 58(12)   &  5265 & $            $  & $             $& (23/2)     & $\rightarrow$ &  21/2      \\
  934.0	 & 75(12)   &  8196 & $  0.253(54) $  & $ -0.183(81)  $& (39/2)     & $\rightarrow$ & (35/2)     \\
  945.8	 & 57(8)    &  3739 & $ -0.16(11)  $  & $  0.19(16)   $&  23/2      & $\rightarrow$ & $21/2^{+}$ \\
  1000.4 & 234(18)  &  1001 & $ -0.050(36) $  & $ -0.076(53)  $& $11/2^{+}$ & $\rightarrow$ & $ 9/2^{+}$ \\
  1032.8 & 66(14)   &  2887 & $  0.16(10)  $  & $  0.13(16)   $& $19/2^{-}$ & $\rightarrow$ & $17/2^{+}$ \\
  1042.8 & 43(6)    &  4782 & $  0.148(81) $  & $ -0.13(13)   $&  27/2      & $\rightarrow$ &  23/2      \\
  1055.0 & 68(12)   &  9251 & $  0.241(24) $  & $ -0.089(37)  $& (43/2)     & $\rightarrow$ & (39/2)     \\
  1143.8 & 35(5)    &  5180 & $  0.248(57) $  & $ -0.038(85)  $& $29/2^{-}$ & $\rightarrow$ & $25/2^{-}$ \\
%
  1220.1 & 49(12)   & 10471 & $  0.297(41) $  & $ -0.045(61)  $& (47/2)     & $\rightarrow$ & (43/2)     \\
  1278.0 & 45(5)    &  3280 & $  0.327(39) $  & $ -0.067(55)  $& $19/2^{-}$ & $\rightarrow$ & $19/2^{+}$ \\
  1333.2 & 50(4)    &  6983 & $ -0.319(29) $  & $  0.059(44)  $& $33/2^{-}$ & $\rightarrow$ & $31/2^{-}$ \\
  1389.0 & 30(8)    & 11860 & $  0.263(29) $  & $  0.028(45)  $& (51/2)     & $\rightarrow$ & (47/2)     \\
  1414.0 & 690(60)  &  1414 & $  0.047(29) $  & $  0.004(43)  $& $13/2^{+}$ & $\rightarrow$ & $ 9/2^{+}$ \\
  1428.1 & 199(18)  &  3280 & $ -0.137(25) $  & $  0.084(38)  $& $19/2^{-}$ & $\rightarrow$ & $17/2^{+}$ \\
  1437.4 & 290(18)  &  3439 & $ -0.207(22) $  & $  0.049(35)  $& $21/2^{-}$ & $\rightarrow$ & $19/2^{+}$ \\
  1527.0 & 18(4)    &  5265 &                 &                & (23/2)     & $\rightarrow$ & 23/2       \\
  1532.3 & 172(14)  &  3534 & $ -0.337(10) $  & $ -0.036(15)  $& $21/2^{-}$ & $\rightarrow$ & $19/2^{+}$ \\
  1573.3 & 20(2)    & 13434 & $  0.29(17)  $  & $ -0.06(27)   $& (55/2)     & $\rightarrow$ & (51/2)     \\
  1746.5 & 48(6)    &  3749 & $ -0.322(67) $  & $  0.026(96)  $& $21/2^{-}$ & $\rightarrow$ & $19/2^{+}$ \\
  1789.4 & 9(2)     & 15223 & $            $  & $             $& (59/2)     & $\rightarrow$ & (55/2)     \\
  1969.7 & 3(1)     & 17193 & $            $  & $             $& (63/2)     & $\rightarrow$ & (59/2)     \\
  2330.5 & 17(2)    &  4333 & $  0.23(9)   $  & $  0.24(13)   $&  21/2      & $\rightarrow$ & $19/2^{+}$ \\
\hline
\hline
\end{longtable*}

\begin{figure}[ht]
  \includegraphics[width=8.6cm]{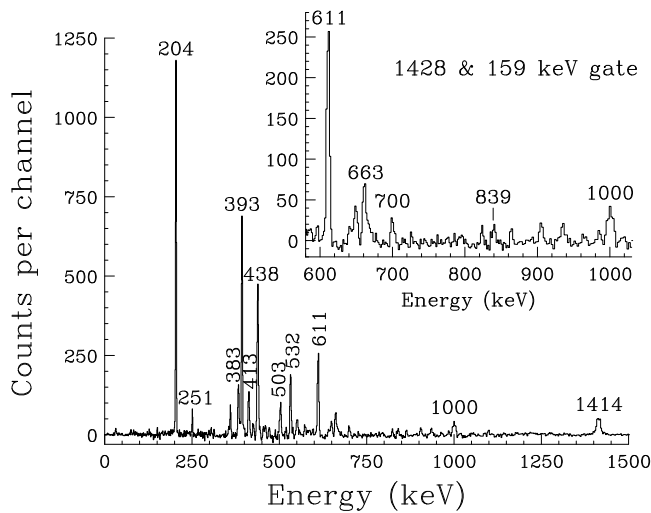}
\caption{%
Gamma-ray energy spectrum gated by the 1428 and 159~keV transitions.
The inset enlarges the region between 600~keV and 1~MeV.
}
\label{gate1}
\end{figure}

\begin{figure}[ht]
  \includegraphics[width=8.6cm]{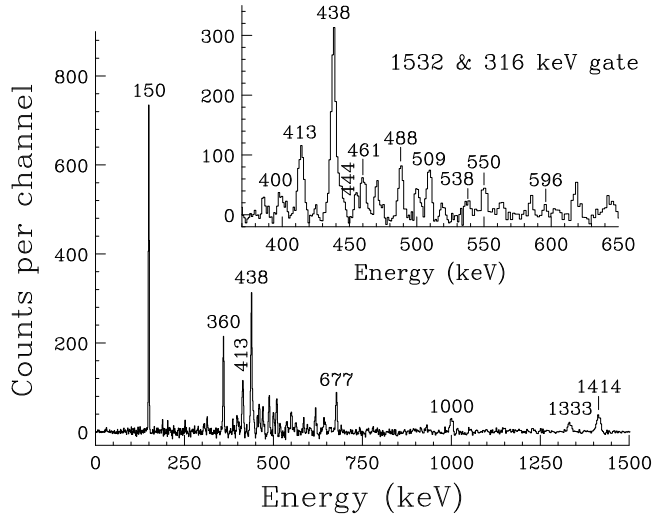}
\caption{%
Gamma-ray energy spectrum gated by the 1532 and 316~keV transitions.
The inset enlarges the region between 350 and 650~keV.
}
\label{gate2}
\end{figure}

%
%

Above the isomeric 19/2$^+$ level, high-energy $\gamma$-ray transitions 
of $\sim$1.5~MeV connect to the 19/2$^{-}$ level at 3.280~MeV and 
21/2$^{-}$ level at 3.534~MeV. 
These levels are suggested to be based on the 
$\pi$g$_{9/2}^{-1}\otimes\nu$(h$_{11/2}$,g$_{7/2}$) configuration, 
i.e. a neutron excitation between the g$_{7/2}$ and h$_{11/2}$ sub-shells.
The two sequences of magnetic dipole transitions have previously been 
observed to connect negative parity states up to 33/2$^{(-)}$ 
\cite{in107_prc58}.

Gamma-ray coincidence relations of previously reported transitions 
\cite{in107_prc58} were confirmed as shown in the right part of partial 
level scheme in Fig.~\ref{level_scheme}.
Table~\ref{gamma_table} summarizes $\gamma$ rays assigned to $^{107}$In. 
%
%
The relative intensity of each $\gamma$ transition was extracted by fitting 
the $\gamma$-$\gamma$ correlation matrix by taking into account the 
$\gamma$-ray intensities of feeding transitions, branching ratio, 
and conversion coefficient of the lower transitions using the ESCL8R program 
in the RADWARE software package \cite{radware}.

%
%
By gating on the E$_\gamma$-E$_\gamma$ matrix and 
the E$_\gamma$-E$_\gamma$-E$_\gamma$ cube, coincidence relations 
between known $\gamma$ rays and newly observed $\gamma$ rays were 
examined.
Fig.~\ref{gate1} shows a $\gamma$ ray energy spectrum obtained by 
double gating in the cube on the 1428 and 159~keV transitions 
which are members of one of the sequences of magnetic dipole transitions 
(M1-band~1). 
Gamma-ray transitions up to the 33/2$^{-}$ level at 6.066~MeV 
(see Fig.~\ref{level_scheme}) were confirmed and two new transitions, 
700 and 839~keV, were placed above this level.
These two transitions also have M1 character as shown in table~\ref{gamma_table}. 

%
%
Fig.~\ref{gate2} shows a spectrum obtained by double-gating in the cube 
on the 1532 and 316~keV transitions which are members of another 
M1~sequence (M1-band~2). 
Gamma-ray transitions up to the 33/2$^{-}$ level at 6.983~MeV were confirmed and 
in addition four $\gamma$-ray transitions were observed in coincidence with the
sequence. 
Based on the angular distribution analysis, the multipolarities of these 
four transitions were assigned as M1 and accordingly the spins and parities 
of the four levels above the 33/2$^{-}$ level were assigned up to the 41/2$^{-}$ 
level at 8.977~MeV.

%
%
\begin{figure}
  \includegraphics[width=8.6cm]{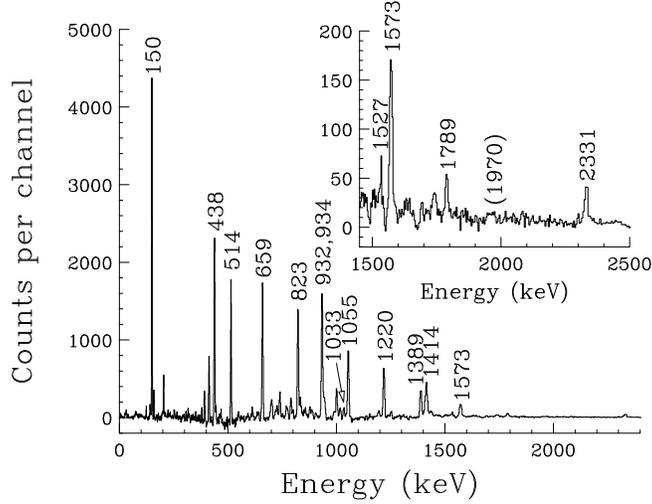}
\caption{%
Sum of $\gamma$-ray energy spectra created by double-gating on 
the in-band transitions of the rotational band (STB1).
The inset enlarges the region between 1.45 and 2.5~MeV.
}
\label{band}
\end{figure}
%
%
\begin{figure}[ht]
  \includegraphics[width=8.6cm]{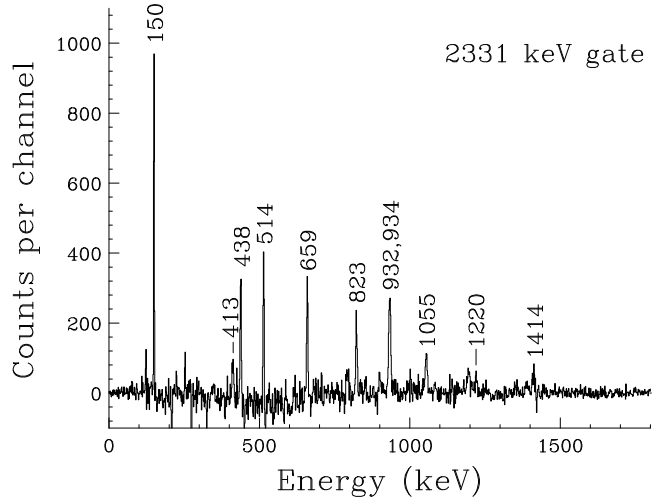}
\caption{%
Gamma-ray energy spectrum gated by the 2331~keV linking transition 
from the $\gamma-\gamma$ matrix.
}
\label{link}
\end{figure}

\begin{figure}[ht]
  \includegraphics[width=8.6cm]{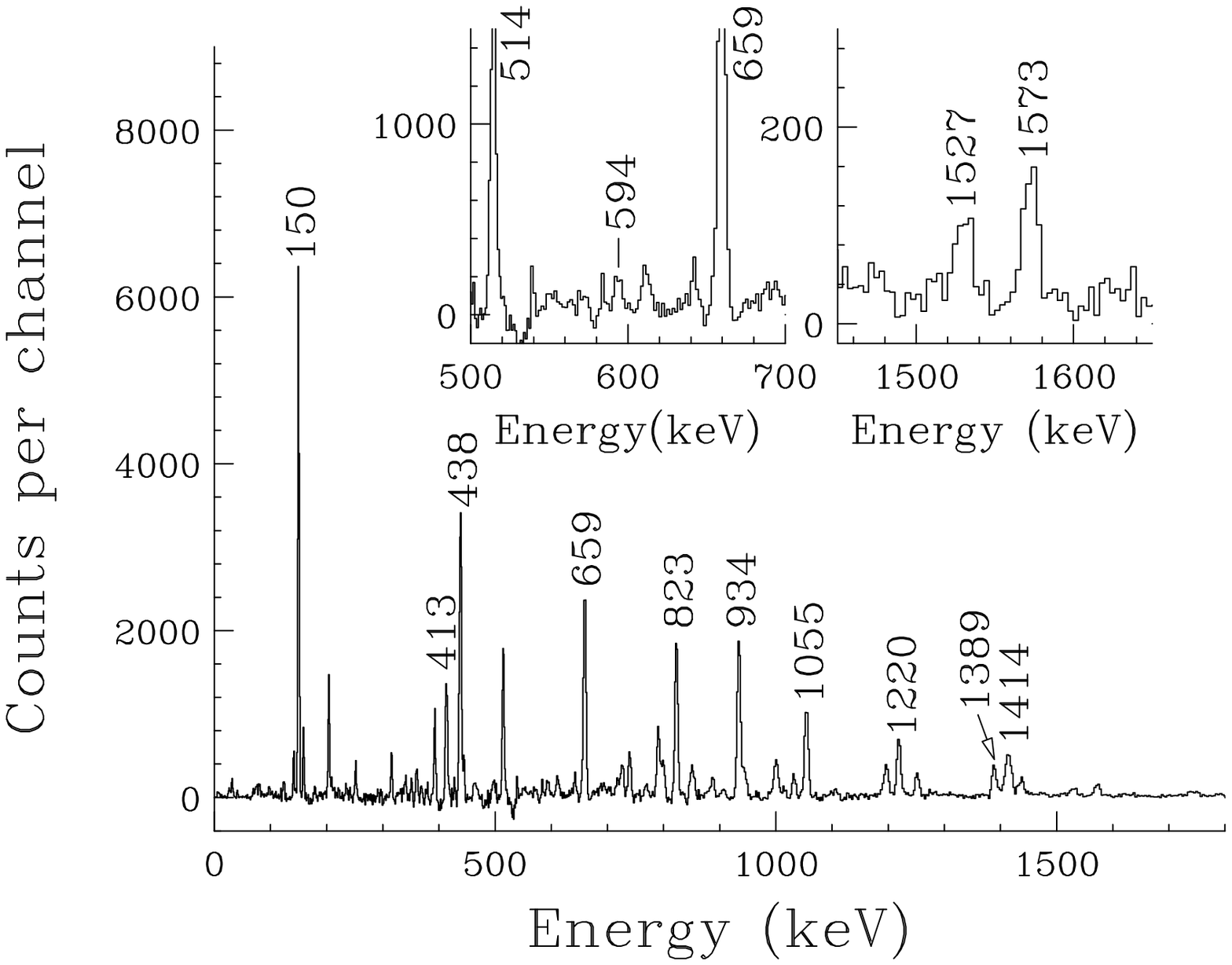}
\caption{%
Gamma-ray energy spectrum created by gating on the in-band transition of the rotational
band in one axis, and the 946, 791, 150, 438, 413, 1000, and 1414~keV transitions 
in the other axis of $\gamma\gamma\gamma$ cube.
The insets enlarge the regions from 0.5~MeV to 0.7~MeV, and from 1.45~MeV to 1.65~MeV,
respectively.
}
\label{link2}
\end{figure}

%
%

In addition to the above mentioned M1 transitions, a rotational 
$\gamma$-ray cascade (STB1) consisting of the 514, 659, 823, 934, 1055, 
1220, 1389, 1573, and 1789~keV transitions was observed by gating on 
the E$_\gamma$-E$_\gamma$ matrix as shown in Fig.~\ref{level_scheme} 
and \ref{band}. 
A 1970~keV transition was tentatively placed on top of the band.
%
%
The 934~keV transition has a larger intensity than the other neighboring 
in-band transitions, 823 and 1055~keV since it is a self-coincident doublet
decomposed of a 932~keV and a 934~keV transition. 
Based on the intensity balance of the in-band transitions, a rotational band
structure was assigned as shown in the left part of Fig.~\ref{level_scheme}.
%
%
In addition, the 2331~keV $\gamma$ ray is observed to be in coincidence with 
the assigned in-band transitions as well as the low-lying transitions, 
150, 438, 413, and 1414~keV (see Fig.~\ref{link}).
Since this $\gamma$ ray is strongly in coincidence with the lowest-lying 
member of the band, as well as with the 932~keV-934~keV doublet, 
the decay path of the band is formed to precede 
the cascade of 932 and 2331~keV transitions to the 19/2$^+$ state at 
2.002~MeV as shown in the level scheme. 
%
%
Note that the intensity of the 932~keV transition in Table~\ref{gamma_table} 
is larger than the sum of the 2331~keV and 594~keV intensities.  
Our interpretation is that there might be several weak unidentified $\gamma$ 
transitions decaying from the 4.333~MeV level carrying the intensity not 
accounted for.
For example, in the double-gated spectrum on the in-band transitions 
(Fig.~\ref{band}), a peak appeared at 1033~keV. 
This indicates the presence of a linking transition from the 4.333~MeV level 
to the 2.887~MeV level. 
However, any such $\gamma$ transitions were not observed 
due to the low statistics.

There is another decay path of the band via the 1527~keV transition 
from the 5.265~MeV state to the 3.739~MeV state.
Fig.~\ref{link2} shows 
a $\gamma$-ray energy spectrum created by gating on the in-band transitions 
of STB1 on one axis, and the 946, 791, 150, 438, 413, 1000, and 
1414~keV transitions on the other axis of the $\gamma\gamma\gamma$ cube.
As shown in the figure, the 594 and 1527~keV transitions are coincident 
with the in-band transitions above 5.265~MeV level as well as transitions 
below 3.739~MeV level. 
These 594 and 1527~keV transitions correspond to the links from 4.333 to 
3.739~MeV level and from 5.265 to 3.739~MeV level, respectively.
These two decay paths establish the excitation energy of the observed 
lowest level of the band to be 5.265~MeV.

In Fig.~\ref{band}, the 1573~keV peak appears stronger than the 1527~keV peak
compared to the relative intensities presented in Table~\ref{gamma_table}.
Since the spectrum shown in Fig.~\ref{band} was created by double-gating on 
the in-band transitions of STB1 including the 932, 934~keV doublet, it contains
double-gating between the 932~keV $\gamma$ ray and other $\gamma$ rays of STB1. 
Therefore, the counts in the 1573~keV peak in Fig.~5 was artificially increased 
relative to that of the 1527~keV transition. 

\begin{figure}[ht]
  \includegraphics[width=8.6cm]{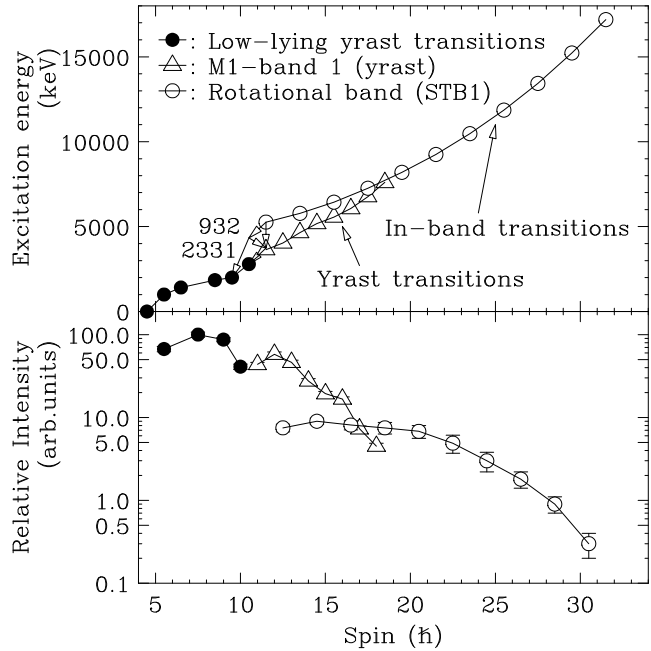}
\caption{%
Excitation energies of the yrast levels and the levels in the rotational 
band as a function of spin (upper panel).
Relative intensities of yrast $\gamma$ transitions and those of the band
are plotted as a function of spin in the lower panel.
}
\label{jex_int}
\end{figure}

The excitation energies of the yrast levels and the levels in the STB1 were 
plotted as a function of spin in the upper panel of Fig.~\ref{jex_int}.
In the lower panel of Fig.~\ref{jex_int}, relative intensities of the yrast 
transitions and the in-band transitions are plotted.  
One can see that the M1-band~1 becomes yrast above spin 10~$\hbar$, 
and that the STB1 becomes yrast around spin 18~$\hbar$.
The intensity profiles are consistent with the trend of yrast sequences formed by 
M1-band~1 and STB1.

%
\begin{figure}[ht]
  \includegraphics[width=8.6cm]{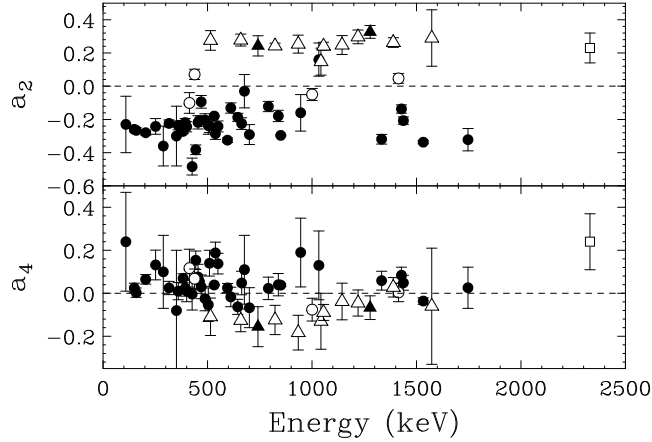}
\caption{%
A plot of a$_{2}$ and a$_{4}$ coefficients of $\gamma$-ray angular distributions, 
W($\theta$)=a$_{0}$(1+a$_{2}$P$_{2}$(cos$\theta$)+a$_{4}$P$_{4}$(cos$\theta$)), 
as a function of $\gamma$-ray energy.
Here, $\theta$ is the emission angle of $\gamma$ rays relative to the beam axis and 
P$_{n}$(cos$\theta$) are the Legendre polynomials.
Filled circles and filled triangles correspond to the angular distribution of 
known M1 and $\Delta$I=0 E1 transitions, respectively. 
Open circles indicate those of the transitions below the isomeric state at 
1.852~MeV.
Open triangles correspond to those of the identified in-band transitions 
as well as known E2 transition (1144~keV).
The open square is for the case of the 2331~keV linking transition.
}
\label{a2a4}
\end{figure}

%
%
Multipolarity assignments for the observed $\gamma$-ray transitions were deduced 
from the angular distribution analysis.
Fig.~\ref{a2a4} displays the a$_{2}$ and a$_{4}$ coefficients of 
the $\gamma$-ray angular distribution function plotted as a function of 
the $\gamma$-ray energy.
Filled circles and filled triangles correspond to the angular distribution of 
known M1 transitions and $\Delta$I=0 E1 transitions, respectively, which agree 
with theoretical estimates.
Open circles are for the $\gamma$-ray transitions decaying from the isomeric
state at 1.852~MeV and they show small values as expected.
Open triangles correspond to those of the newly identified cascade transitions 
which are consistent with a stretched 
E2 character.

\begin{figure}[ht]
  \includegraphics[width=8.6cm]{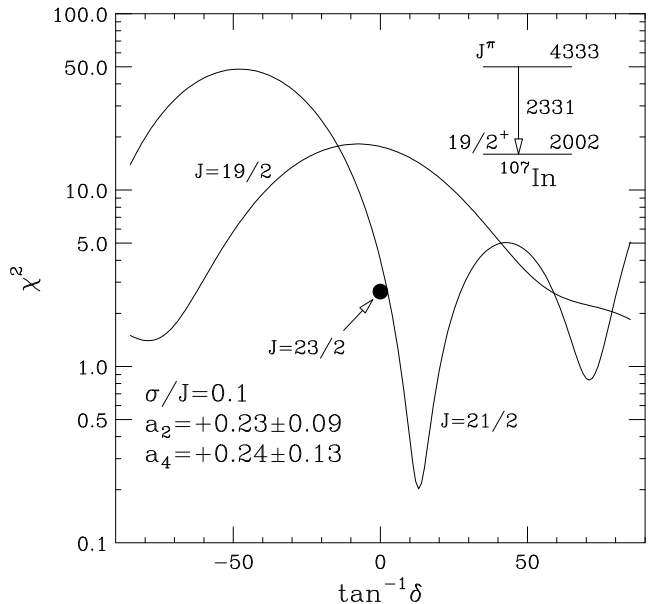}
\caption{%
$\chi^{2}$ vs tan$^{-1}\delta$ curves for the 2331~keV transition in $^{107}$In.
The parameters indicated were used to compute the curves.
}
\label{delta2331}
\end{figure}

%
%
As shown in Fig.~\ref{a2a4}, the angular distribution of the 2331~keV linking
transition (open square) does not clearly determine its multipolarity due to 
the low statistics. 
The positive a$_{2}$ value may indicate that it contains a quadrupole component 
in its multipolarity and the positive a$_{4}$ may reflect a mixed transition. 
In order to try and resolve this ambiguity, the mixing ratio that satisfies the 
99.9$\%$ confidence limit test in $\chi^{2}$ (defined as in equatoin (\ref{eq1}) 
\cite{ang_delta}) is plotted as a function of tan$^{-1}\delta$ in Fig.~\ref{delta2331}.
Three different hypotheses were tested, 19/2$\rightarrow$19/2, 
21/2$\rightarrow$19/2, and 23/2$\rightarrow$19/2.
\begin{equation}
\chi^{2} = \frac{[a_{2} - a_{2}(cal)]^{2}}{3(\Delta a_{2})^{2}}
          + \frac{[a_{4} - a_{4}(cal)]^{2}}{3(\Delta a_{4})^{2}}
\label{eq1}
\end{equation}
As indicated in Fig.~\ref{delta2331}, a parameter '$\sigma$/J' is used to 
implement the initial alignment, where a Gaussian distribution 
for the magnetic substates is assumed and $\sigma$ and J denote the width and 
the spin, respectively. In the calculation, a ratio $\sigma$/J = 0.1 was assumed.
The $\chi^{2}$ result suggests a mixed E2/M1 or M2/E1 transition, 
21/2$\rightarrow$19/2$^{+}$ with a mixing ratio of 0.23. 
Since Weisskopf estimate of lifetimes for a 2331~keV $\gamma$ ray are 
0.4~ps for an E2 and 29~ps for an M2 transition, respectively,
both cases are short enough for prompt $\gamma$ coincidence and
therefore the parity of the 4.333~MeV state was not determined.
Since the decay-out 932~keV $\gamma$-ray transition in cascade with 
the linking transition is close in energy to the 934~keV in-band transition, 
its multipolarity was not firmly identified. 
The statistics of another 1527~keV linking transition were too low to perform
angular distribution analysis in order to establish the multipolarity. 
Under the assumption 
that a dipole transition connects the lowest level of the band with the 
4.333~MeV level, the spin of the band was assigned as starting with (23/2) as
shown in Fig.~\ref{level_scheme}.

\begin{figure}[ht]
  \includegraphics[width=8.6cm]{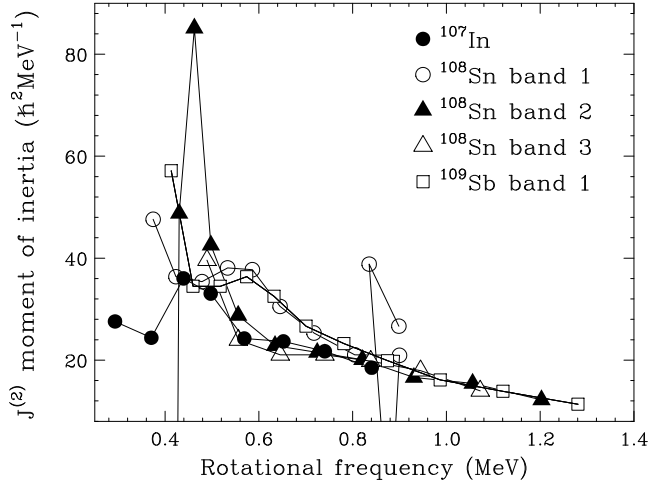}
\caption{%
Experimentally deduced dynamical moments of inertia J$^{(2)}$ in $^{107}$In,
$^{108}$Sn, and $^{109}$Sb. 
The data on  $^{107}$In is from the present work.
}
\label{fig_j2}
\end{figure}

\section{Discussion}
%
%
In this mass region, the presence of rotational bands has been reported previously 
in the N$=$58 isotones, $^{105}$Ag, $^{106}$Cd, $^{108}$Sn, $^{109}$Sb, 
$^{110}$Te, and $^{111}$I \cite{ag105,cd106,sn108,sb109,bob_prl,sb109_2,Te110,i111}.
Among them, rotational bands exhibiting a character of smooth band termination
appeared only in $^{108}$Sn, $^{109}$Sb, and $^{110}$Te.

%
%
  In Fig. \ref{fig_j2}, deduced dynamical moments of inertia J$^{(2)}$
are shown as a function of rotational frequency. 
%
%
Here, the rotational frequency ($\omega$) and the dynamical moment of inertia 
J$^{(2)}$ are deduced from the experimental data (E$_{\gamma}$) using the formulae,
\begin{equation}
\hbar\omega = \frac{dE}{d{\rm I}} \approx \frac{E_{\gamma}}{2}
\end{equation}
and
\begin{equation}
{\rm J}^{(2)} = \frac{d{\rm I}}{d\omega} \approx \frac{4}{\Delta {\rm E}_{\gamma}} ,
\end{equation}
respectively.
In addition to the band in $^{107}$In, also the bands in $^{108}$Sn \cite{sn108}, 
and $^{109}$Sb \cite{sb109_2} are shown, which have positive parity for the neutron
configuration according to the assignments based on the cranked Nilsson-Strutinsky 
calculations \cite{AR-95}.
In general, all J$^{(2)}$ values show the smooth decrease with increasing 
rotational frequency which is characteristic for smooth band termination in 
systems with weak pairing \cite{AR-95,Smooth-PR}. 
This feature is disturbed only at the top of the band~1 in $^{108}$Sn. 
The J$^{(2)}$ values of $^{107}$In show very close similarity with those in bands 
2 and 3 of $^{108}$Sn, while those in band 1 of $^{108}$Sn and band~1 of $^{109}$Sb 
are also very similar. 
This suggests that the structure of the band in $^{107}$In is similar to that 
of bands 2 and 3 in $^{108}$Sn. 
Bands 2 and 3 of $^{108}$Sn were interpreted to be signature partner bands with 
the proton configuration of $\pi[$g$_{9/2}^{-2}\otimes($g$_{7/2}h_{11/2})]$
\cite{sn108, sn108_2}.
On the other hand, band 1 of $^{108}$Sn and $^{109}$Sb were understood to have
configurations of 
$\pi[$g$_{9/2}^{-2}\otimes$g$_{7/2}^2]$ and
$\pi[$g$_{9/2}^{-2}\otimes$g$_{7/2}^2h_{11/2}]$, respectively. 
The hump observed in the J$^{(2)}$ plot for band~1 of $^{108}$Sn and $^{109}$Sb, 
appearing at $\sim$0.55~MeV, was interpreted  as due to the alignment of g$_{7/2}$ 
protons based on cranked Woods-Saxon calculations \cite{sn108_2}.
These results indicate that deformation-driving down-sloping  $\pi$g$_{7/2}$ and 
$\pi$h$_{11/2}$ orbitals and, especially, the holes in the $\pi g_{9/2}$ extruder 
orbitals play an important role to create the rotational band structures which 
undergo smooth band termination in this mass region \cite{AR-95}.
In the case of $^{107}$In, possible configurations of the observed
band (STB1) would be $\pi[$g$_{9/2}^{-2}\otimes$g$_{7/2}]$ or 
$\pi[$g$_{9/2}^{-2}\otimes$h$_{11/2}]$ 
generated by removing one proton from $^{108}$Sn.

\begin{figure}[ht!]
  \includegraphics[width=8.6cm]{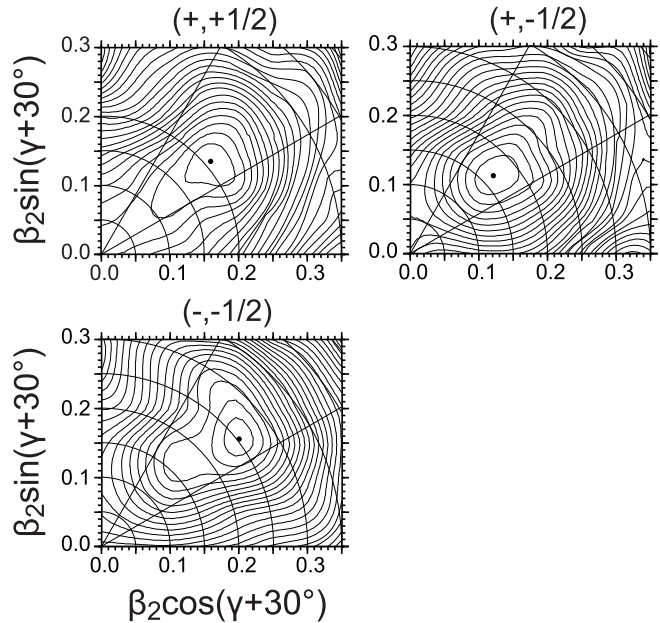}
\caption{%
Calculated total Routhian surfaces (TRS) at a rotational frequency of 
0.75~MeV. 
The lowest positive-parity configurations with signature +1/2 and -1/2 
are shown in the left and right upper panels, respectively, 
while the lowest negative-parity configuration with signature -1/2 is 
plotted on the lower left panel.
}
\label{trs}
\end{figure}

Cranked Strutinsky calculations based on the Woods-Saxon (WS) 
\cite{WoodsSaxon} and Nilsson (CNS) \cite{AR-95,Smooth-PR} potentials 
were performed in order to interpret the structure of the observed 
rotational band (STB1). 
In the former total Routhian Surface (TRS) calculations, 
pairing correlations were taken into account 
by means of a seniority and double stretched quadrupole pairing force \cite{pairing}.
Approximate particle number projection was performed via the Lipkin-Nogami
method \cite{LipkinNogami1,LipkinNogami2}. 
Each quasiparticle configuration was blocked self-consistently. 
The energy in the rotating frame of reference was minimized with respect to the 
deformation parameters $\beta_{2}, \beta_{4}$ and $\gamma$.
Deformed minima in the total Routhian surfaces (TRS) were found at ($\beta_2,\gamma$) 
$\sim$ ($0.20,10^{\circ}$) for both signatures at positive parity 
($\pi$,$\alpha$)$=$(+,$\pm$1/2) and at ($\beta_2,\gamma$) $\sim$ ($0.25,8^{\circ}$) 
for the negative parity, negative signature configuration ($-$,$-$1/2) as shown in 
Fig.~\ref{trs}.  
The structure of the experimentally observed rotational band (STB1)  
in $^{107}$In is likely to correspond to one of these configurations.

\begin{figure}[ht]
  \includegraphics[width=8.6cm]{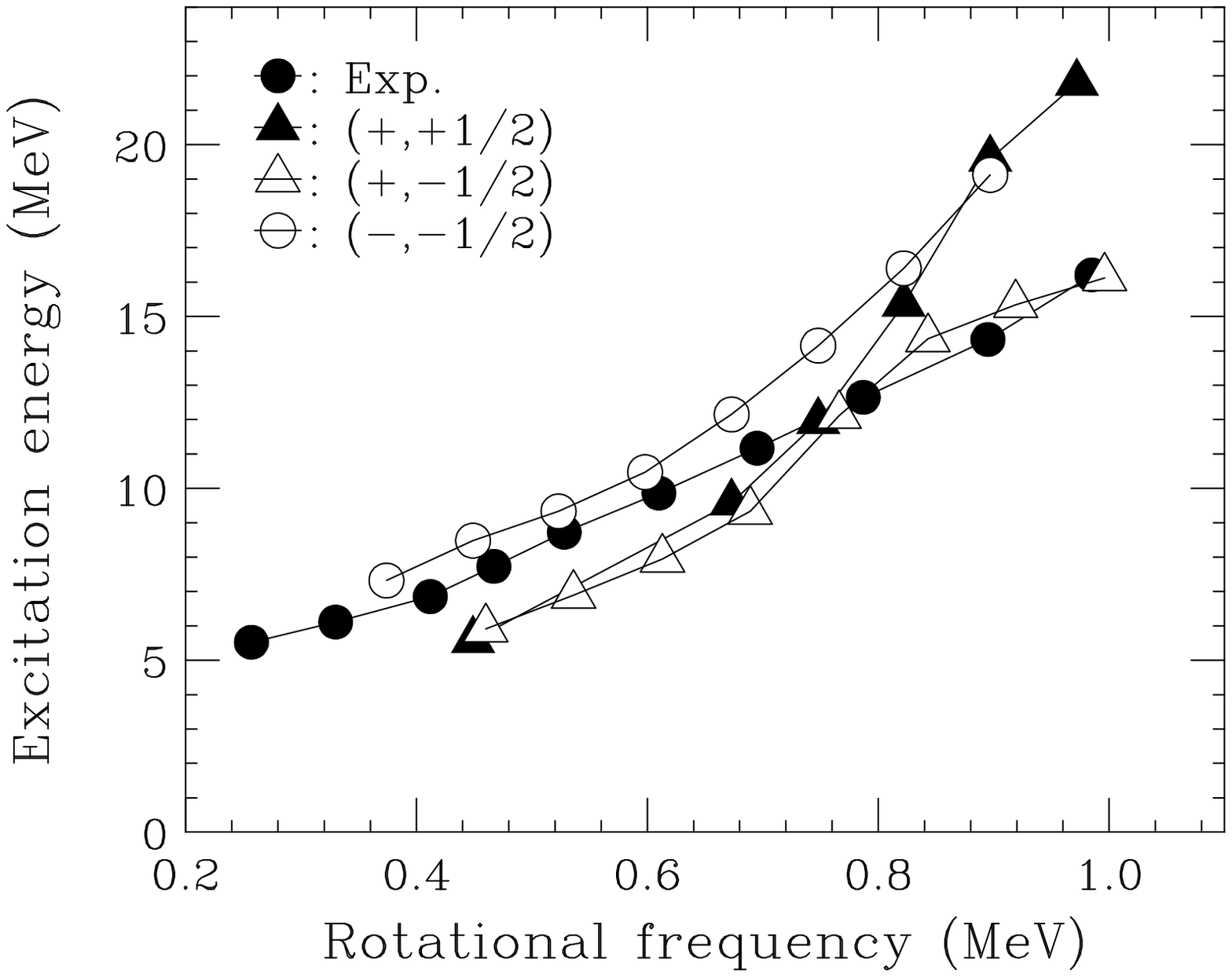}
  \includegraphics[width=8.6cm]{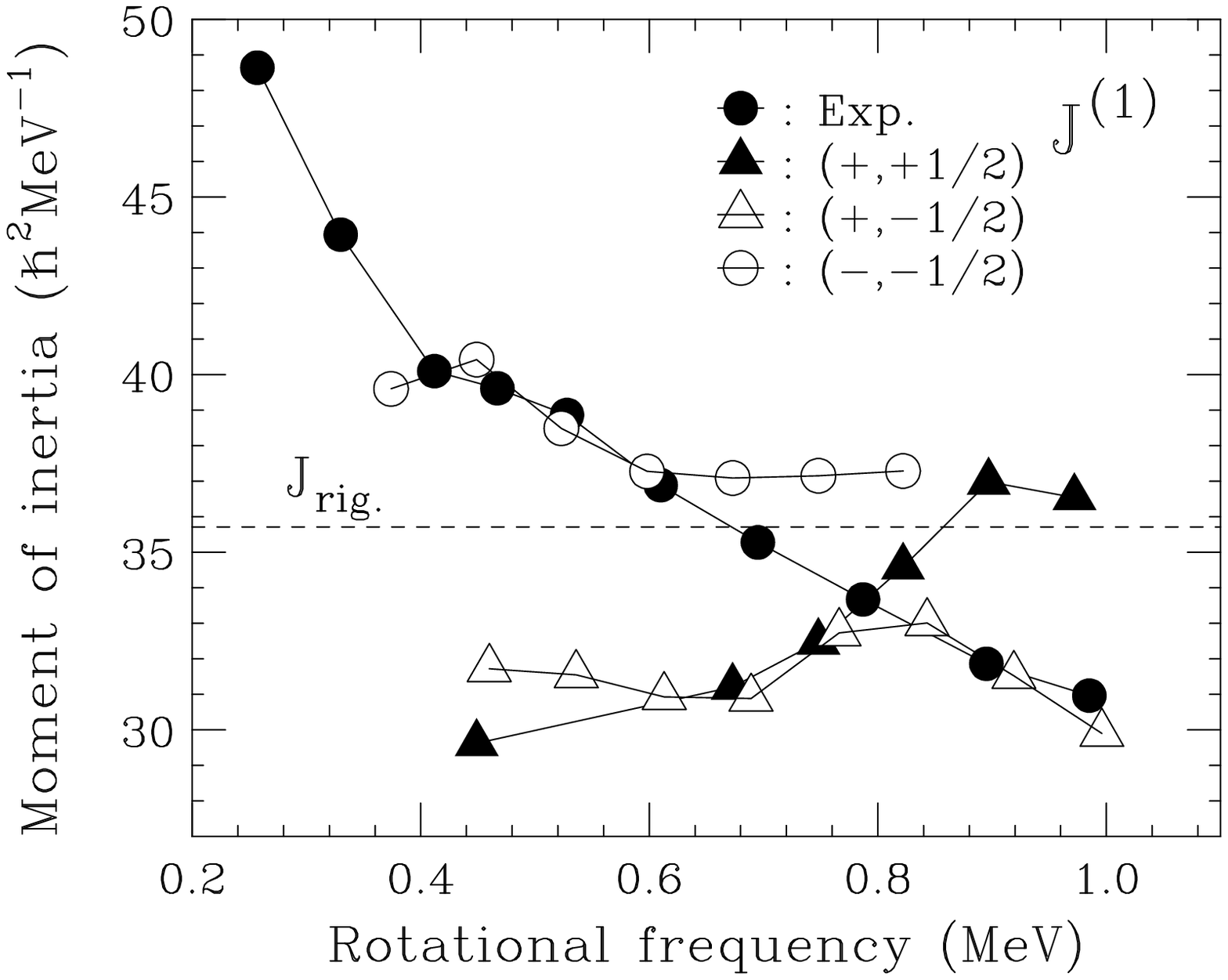}
  \includegraphics[width=8.6cm]{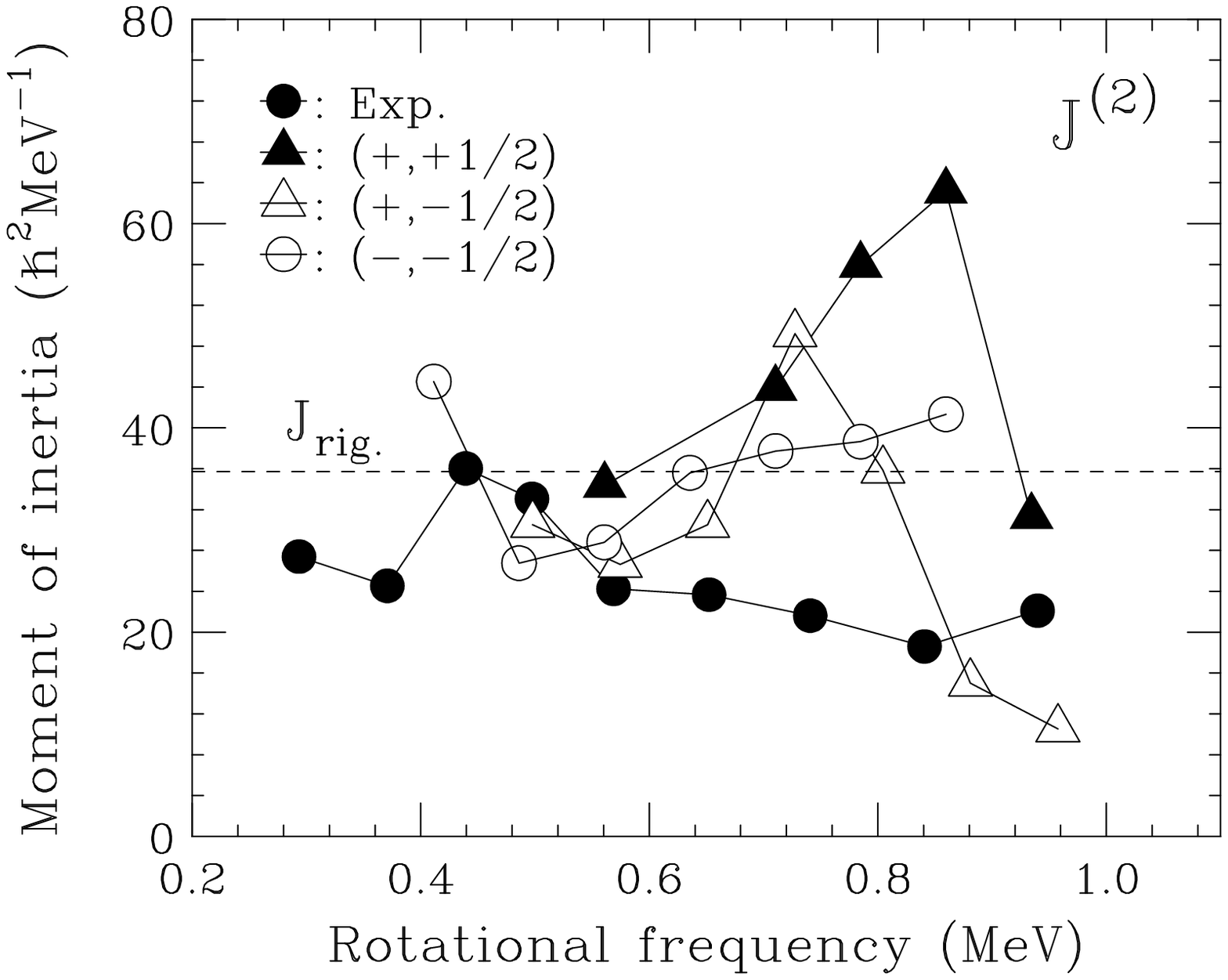}
\caption{
Comparison of experimental and calculated excitation energies, J$^{(1)}$ and 
J$^{(2)}$ moments of inertia as a function of rotational frequency.
The highest-frequency data point, which is originated from the tentatively 
assigned 1970~keV transition, is also included in the plot.
Theoretical values were obtained by TRS calculations.
The dashed line marks the rigid-body moment of inertia for a deformation 
parameter $\beta_{2}=$0.25.
}
\label{trs_j1j2}
\end{figure}

\begin{figure*}[ht!]
\centering
\includegraphics[width=16.0cm,angle=0]{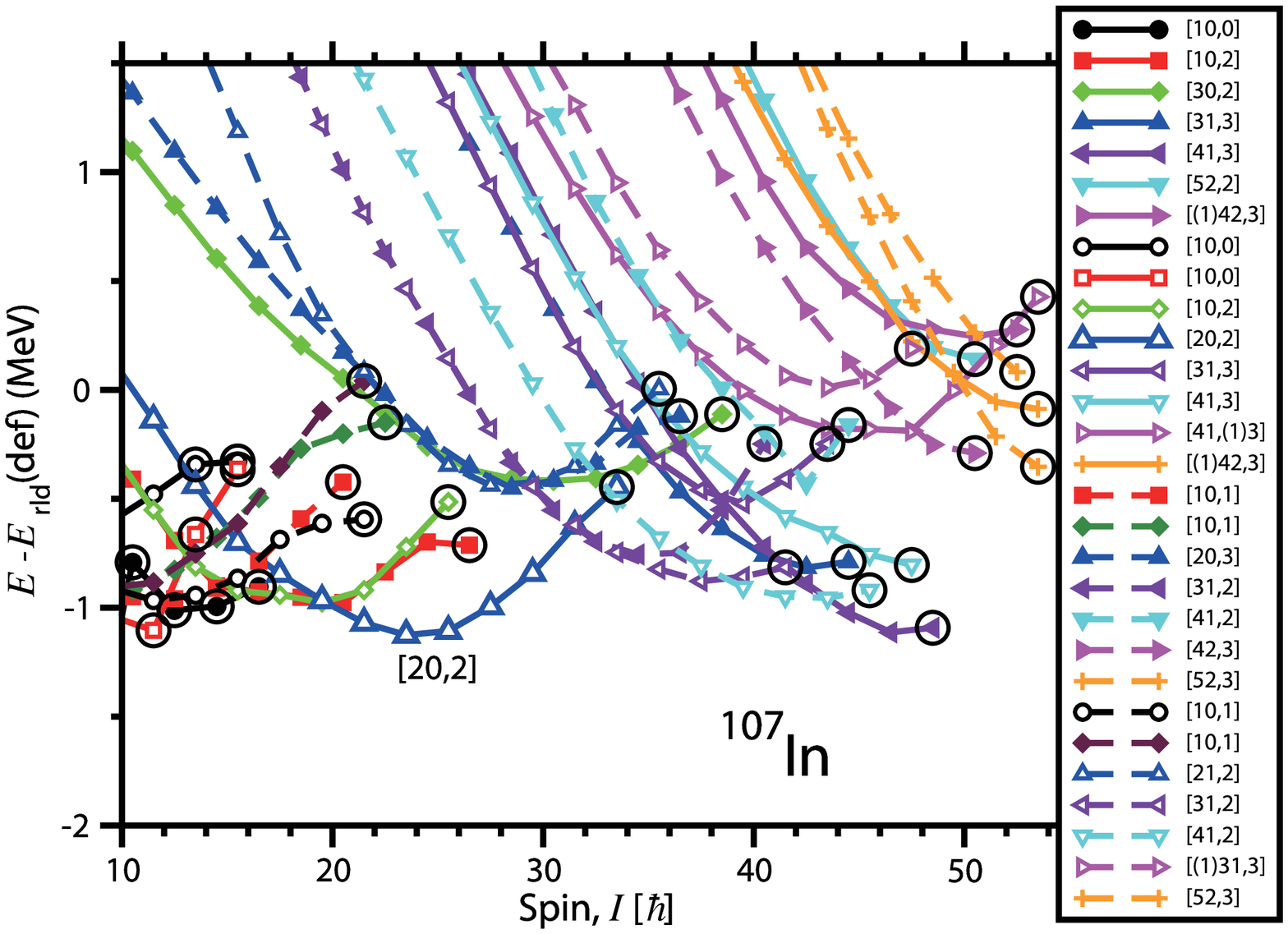}
\vspace{0.5cm}
\caption{%
(Color online) Calculated energies of the configurations forming the yrast 
line in four combinations of parity and signature shown relative to rigid 
rotor reference.  
Large open circles indicate terminating and aligned states. 
Solid and dashed lines are used for positive and negative parity states, 
respectively. 
Solid and open symbols are used for the signatures $\alpha=+1/2$ and 
$\alpha=-1/2$, respectively. 
Configuration labels are shown in the legend.}
\label{In107-eld}
\end{figure*}
%
%
Excitation energies, kinematic (J$^{(1)}$) and dynamic (J$^{(2)}$) moments of 
inertia obtained from the TRS calculations are compared with experimental 
values in Fig.~\ref{trs_j1j2}. 
In order to deduce the experimental J$^{(1)}$-moment of inertia, 
we have assumed the spin of the lowest level of the band to be 23/2. 
The TRS calculation shows that the configuration of positive parity with negative 
signature $(+, -1/2)$ is lowest in energy among the predicted configurations 
in the frequency range of interest.
However, as can be seen in Fig.~\ref{trs_j1j2}, J$^{(1)}$ and J$^{(2)}$ moments of 
inertia for $(+, \pm1/2)$ configurations show quite different behavior comparing 
with the experimental values.
On the other hand, the configuration with $(-,-1/2)$ is in better agreement 
with the experimental properties except for the high frequency region.
However, it is quite likely that the TRS calculations overestimate the role of 
pairing at high spin and this leads to observed discrepancies between experiment 
and calculations. 
%


In the CNS \cite{AR-95,Smooth-PR} calculations, the pairing is neglected and 
thus the results are expected to be realistic only at high spin, 
I$\ge$15$\hbar$. 
In general, a very good agreement \cite{Smooth-PR} between the CNS calculations 
and observed high-spin bands in this mass region has been obtained with the
$\kappa$- and $\mu$-values from Ref.\ \cite{par} and these are the values which 
are used also in the present study.
The calculations in the current analysis are performed using the CNS version of 
Ref.\ \cite{CR.06}, so that not only the relative energies between the bands but 
also the absolute energy scale can be compared.  
The single-particle configurations are defined by the occupation of low- and 
high-$j$ orbitals. 
The total energy of each configuration is minimized at each spin in the 
$(\varepsilon_2, \varepsilon_4, \gamma)$ deformation plane. 
Standard CNS labeling of each configuration is used. 
This shorthand notation is based on the number of particles in different $j$ 
shells for each configuration. 
The $j$ shells are pure only if the shape is spherical. 
Thus, in general the labeling refers to the dominant shell only, while the wave 
functions also contain components from other $N$ and $j$ shells. 
Relative to a closed $^{100}$Sn core, a shorthand configuration label 
$[(p_1)p_2p_3,(n_1)n_2]$ can be written as \cite{Smooth-PR}
\begin{eqnarray}
[(p_1)p_2p_3,(n_1)n_2] & = & \pi (N=3)^{-p_1} (g_{9/2})^{p_2} (h_{11/2})^{p_3} 
\nonumber \\ 
                  &   & \otimes\,\, \nu  (N=3)^{-n_1}(h_{11/2})^{n_2} 
\end{eqnarray}
with the remaining particles (protons/neutrons) outside the core located in 
the mixed $(d_{5/2}\,g_{7/2})$ orbitals. 
The label $p_1$ ($n_1$) is dropped if no holes are generated in the $N=3$ proton 
(neutron) shell.

\begin{figure*}[ht!]
\centering
\includegraphics[width=13cm,angle=0]{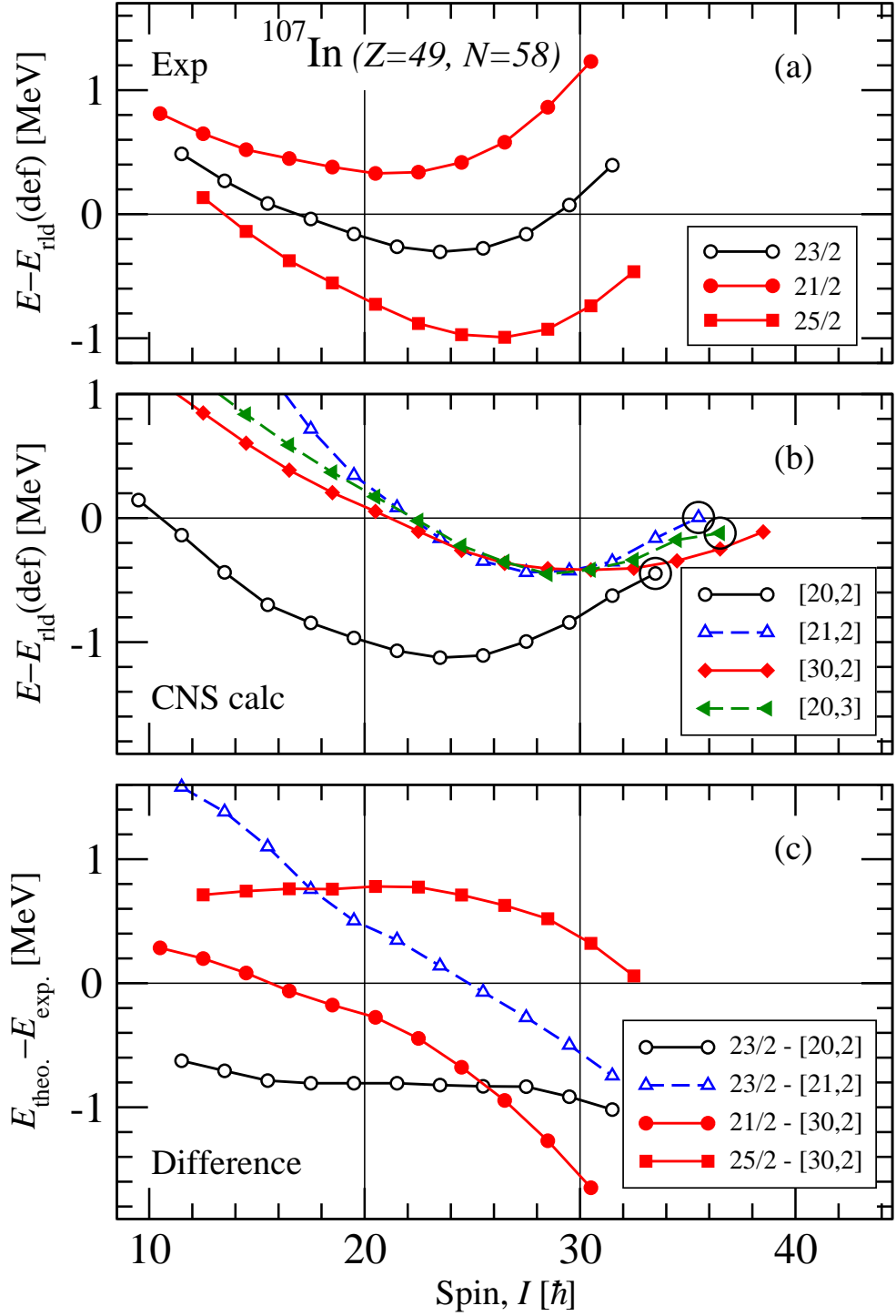}
\caption{%
 (Color online) Comparison between the experimentally observed $\Delta I=2$
rotational band (STB1) and the CNS predictions for selected configurations. 
The top panel shows 
the experimental band with different spin assignments $I_0$ for the lowest 
observed state. 
The middle panel shows the chosen predicted configurations. 
The bottom panel plots the energy difference between the predictions and 
observations.
}
\label{exp-th}
\end{figure*}

The results of the calculations are shown in Fig.\ \ref{In107-eld}. 
One can see that the [20,2] configuration dominates the yrast line in the 
spin range $I=20-31\hbar$. 
This configuration is therefore the most likely counterpart for the observed decoupled 
$\Delta I=2$ band. 
Fig.\ \ref{exp-th} compares the experimental band with the results of the 
calculations. 
Fig.\ \ref{exp-th}a shows the experimental band for three different spin 
assignments $I_0$ for the lowest state in the band. 
Fig.\ \ref{exp-th}b displays the configurations which can be possible 
counterparts of the experimental band. 
In Fig.\ \ref{exp-th}a,b, the $y$ axis denotes the energy with the rotating 
liquid drop (rld) energy \cite{CR.06} subtracted. 
The energy difference between the predictions and observations is plotted in 
Fig.\ \ref{exp-th}c. 
In the ideal case when the transition energies will be predicted correctly by 
the calculations for each transition in the band, the curve of the band in 
Fig.\ \ref{exp-th}c will have a constant energy difference for all states of 
a given configuration, i.e. it will be horizontal. 
Furthermore, if the absolute energy is predicted correctly, this difference 
should be zero with the experience \cite{CR.06} that the difference will 
generally be smaller than one MeV, i.e. the absolute energy at high spin can be 
described with a similar accuracy as ground state masses. 
One can see that reasonable agreement between theory and experiment is obtained 
when the [20,2] configuration is assigned to the band; the deviation from the 
horizontal line in Fig.\ \ref{exp-th}c of $\sim 1$ MeV is at the 
upper limit of the typical discrepancies between theory and experiment. 
For this configuration assignment, the observed  band in $^{107}$In 
is one transition short of termination. 
Further support for this configuration assignment comes 
from the analysis of the configurations of the smooth terminating bands in the $N=58$ 
isotones (see Fig.\ 23 in Ref.\  \cite{Smooth-PR}). 
In $^{108}$Sn, the yrast line in the spin range $I=20-28\hbar$ is dominated 
by the [20,2] configuration. 
This is exactly the same configuration (in terms of shorthand notation) 
as we assign to the $\Delta I=2$ band in $^{107}$In. 
The difference between the [20,2] configurations in these two nuclei is related 
to an occupation of the extra $(d_{5/2}\,$g$_{7/2})$ proton in $^{108}$Sn.
Note also that this configuration assignment fits exactly into the systematics 
of the configurations of smooth terminating bands observed 
in this mass region (see Fig.\ 20 and 23 in Ref.\ \cite{Smooth-PR}). 
Indeed, according to this figure, the present signature $\alpha = -1/2$ [20,2] band 
is predicted to be the most favored smooth terminating band in $^{107}$In.  
%

\begin{figure*}[thb!]
\centering\includegraphics[clip=true,width=0.320\textwidth,angle=0]{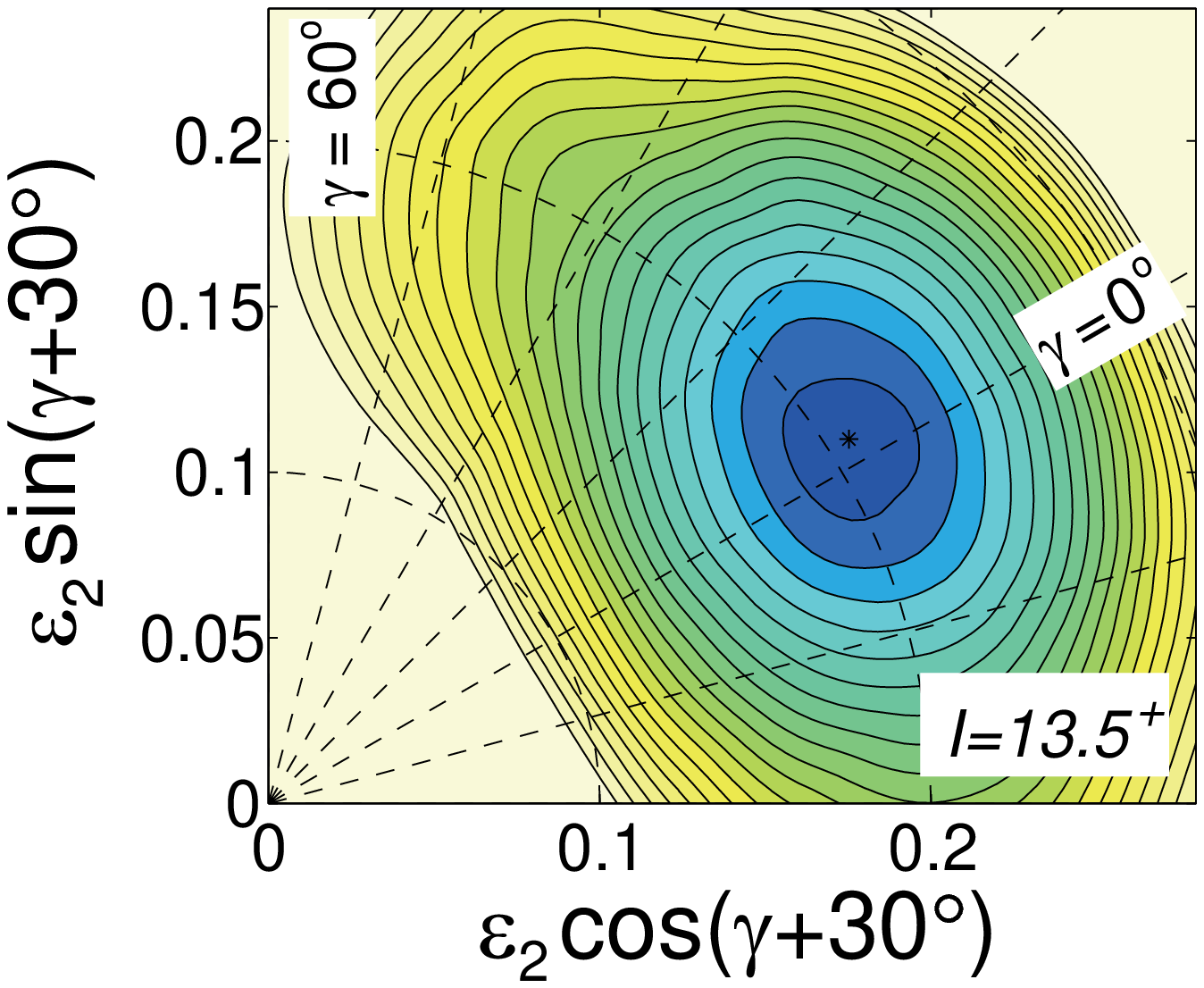}
\centering\includegraphics[clip=true,width=0.320\textwidth,angle=0]{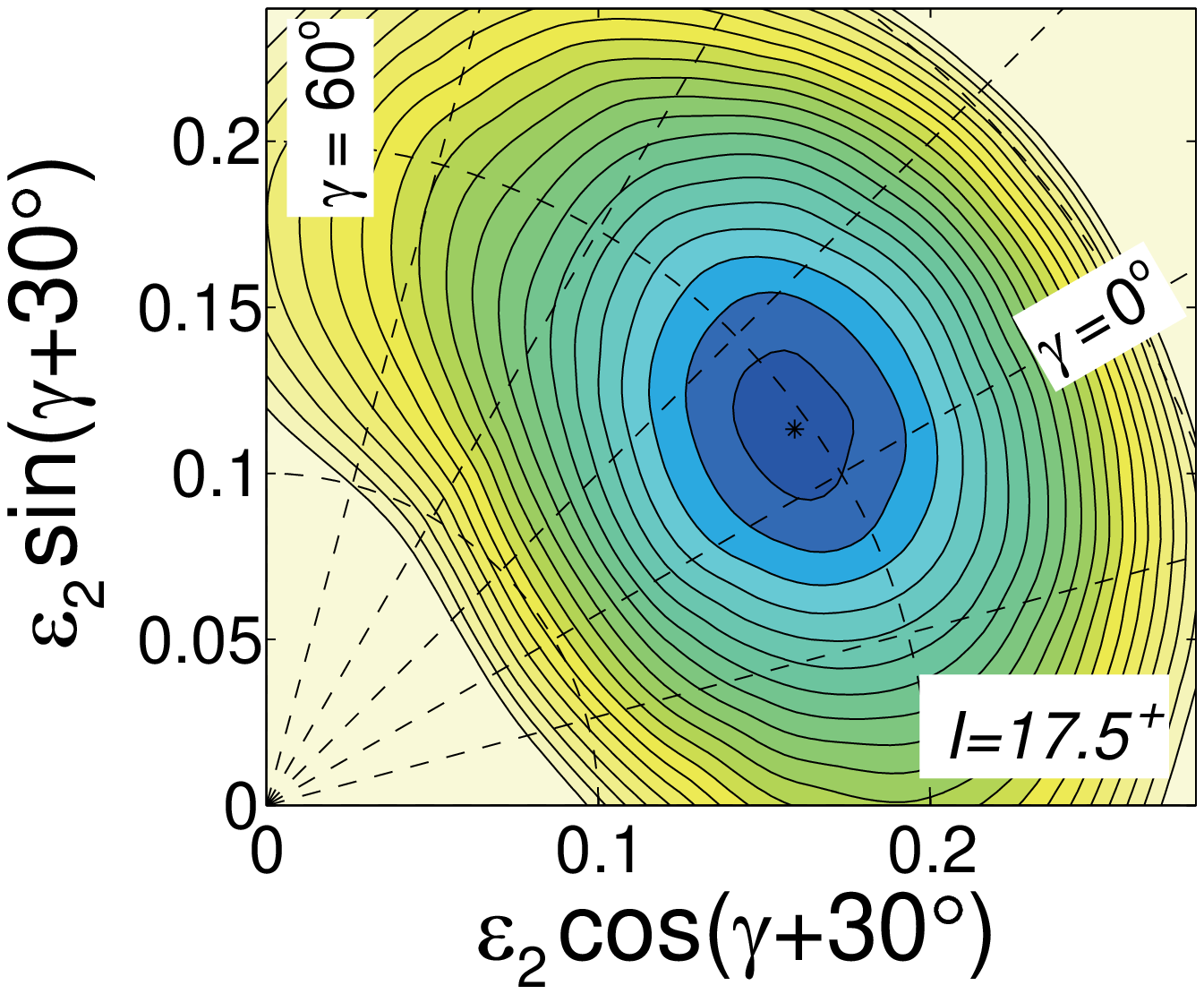}
\centering\includegraphics[clip=true,width=0.320\textwidth,angle=0]{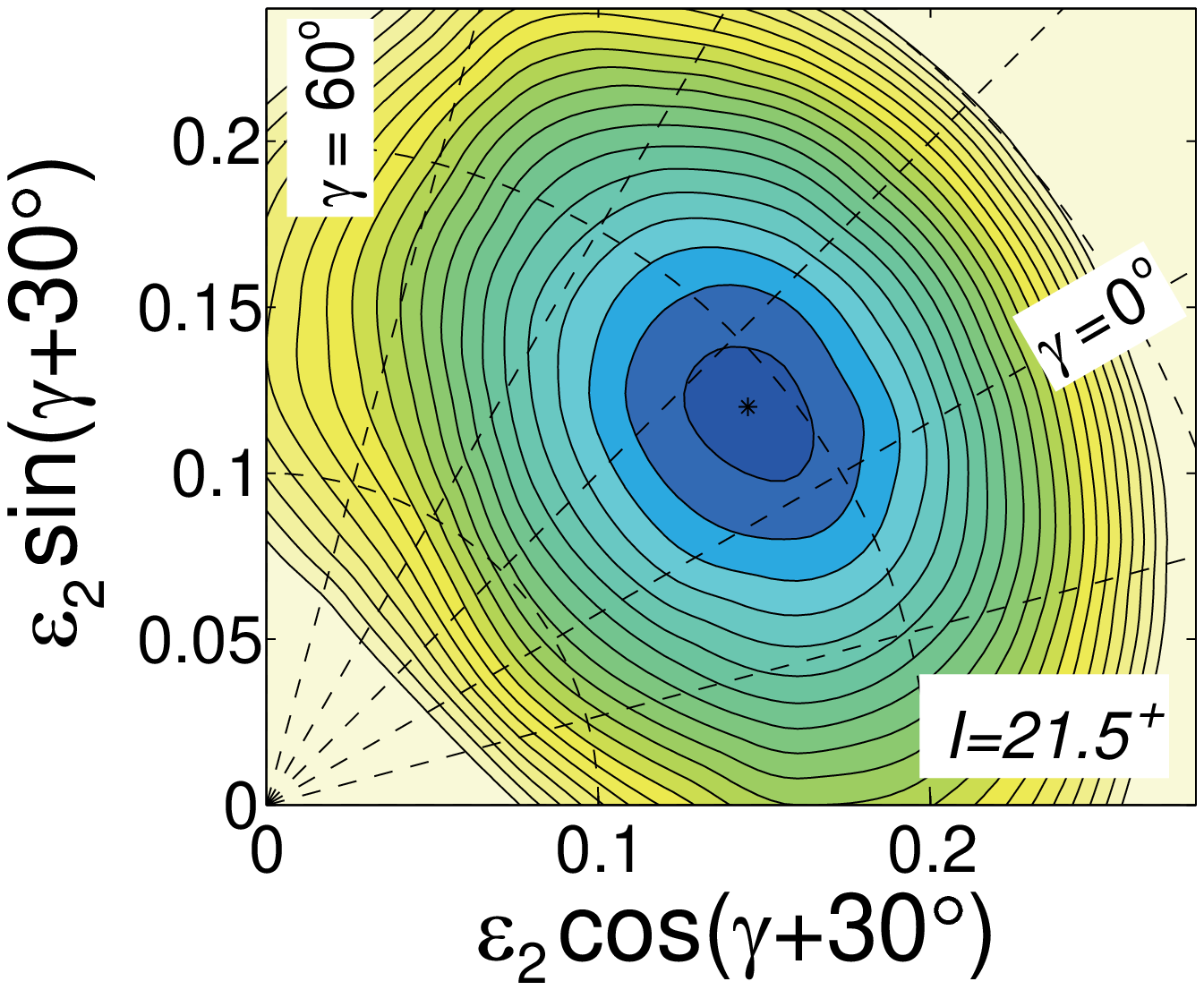}
\centering\includegraphics[clip=true,width=0.320\textwidth,angle=0]{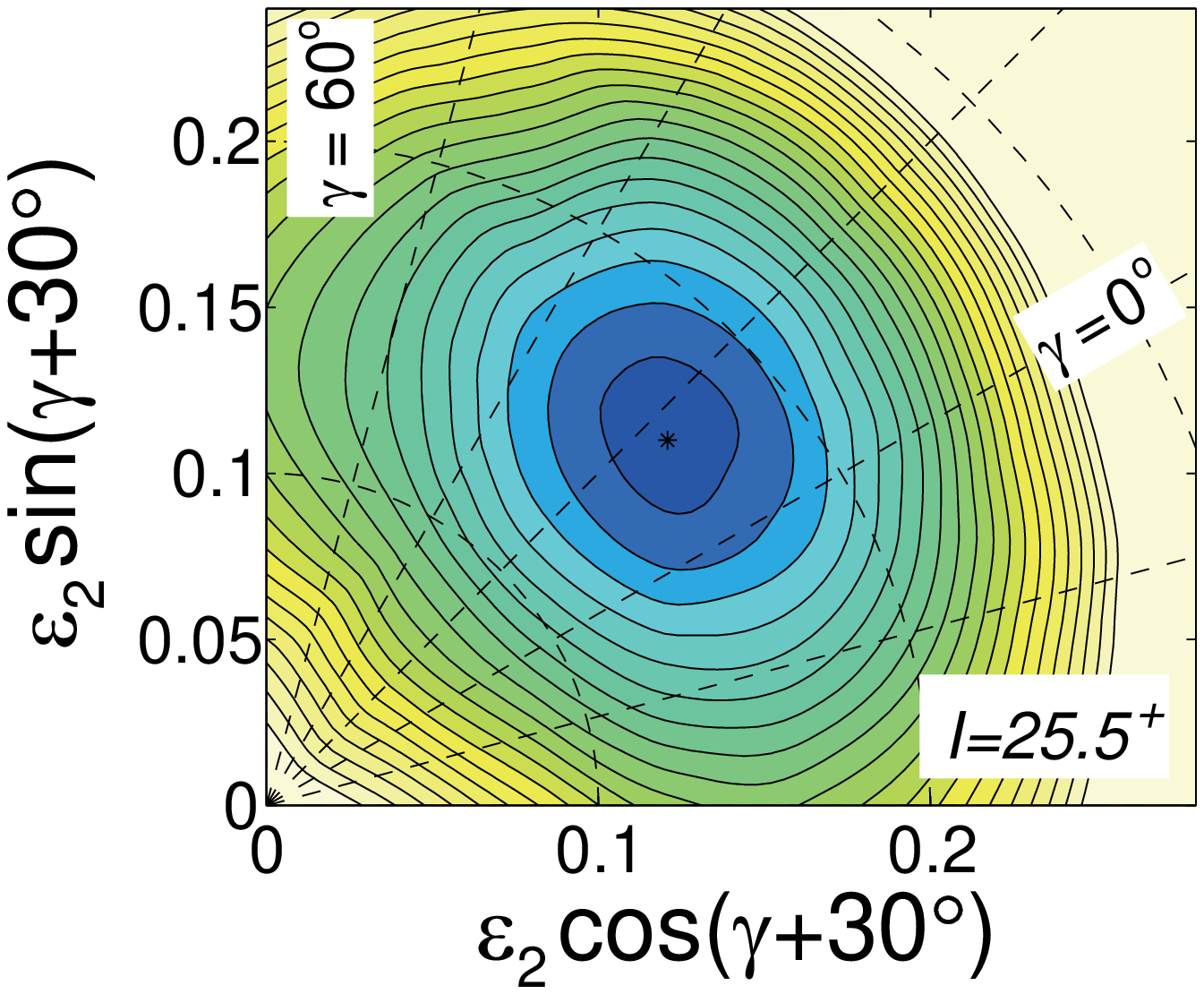}
\centering\includegraphics[clip=true,width=0.320\textwidth,angle=0]{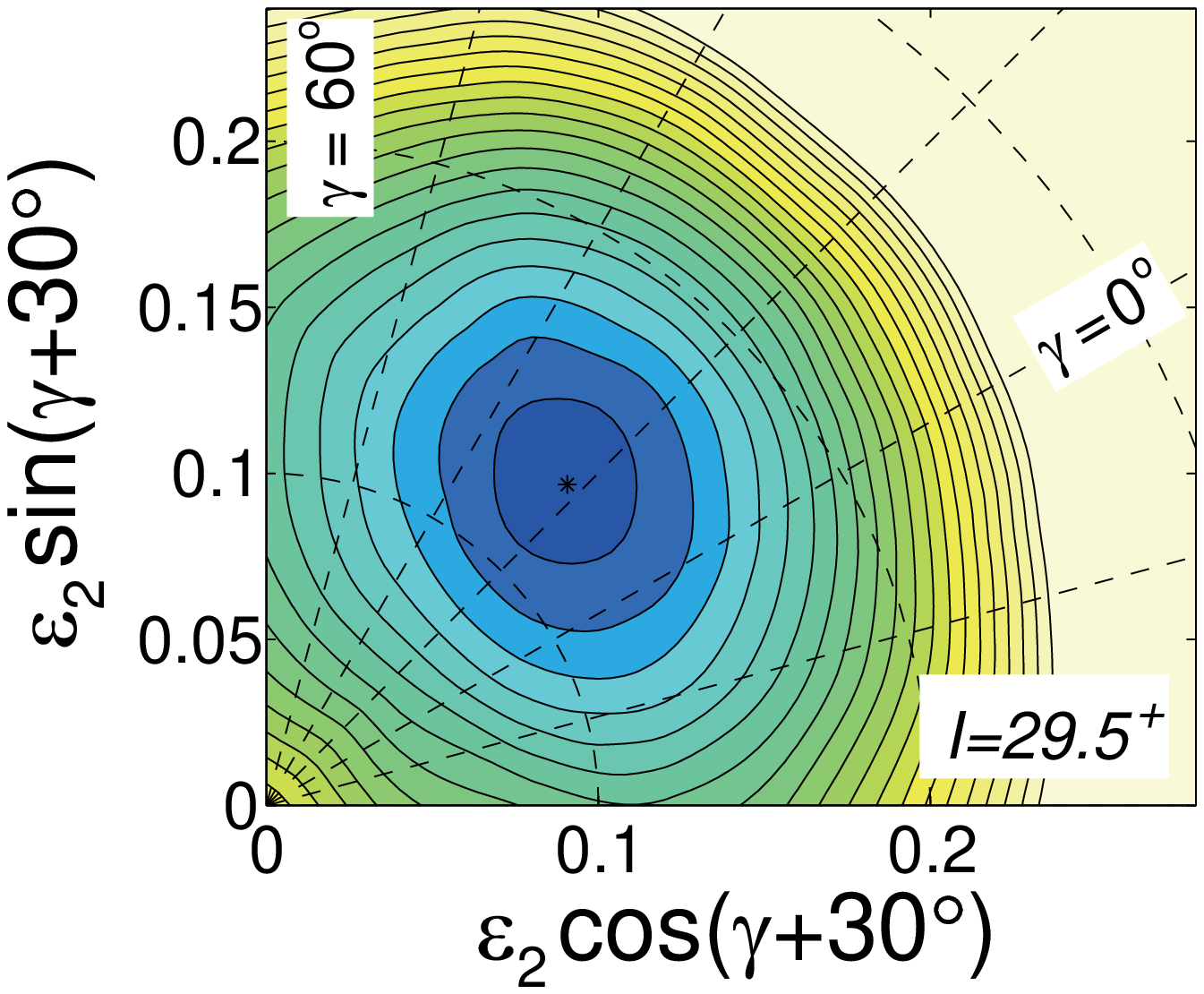}
\centering\includegraphics[clip=true,width=0.320\textwidth,angle=0]{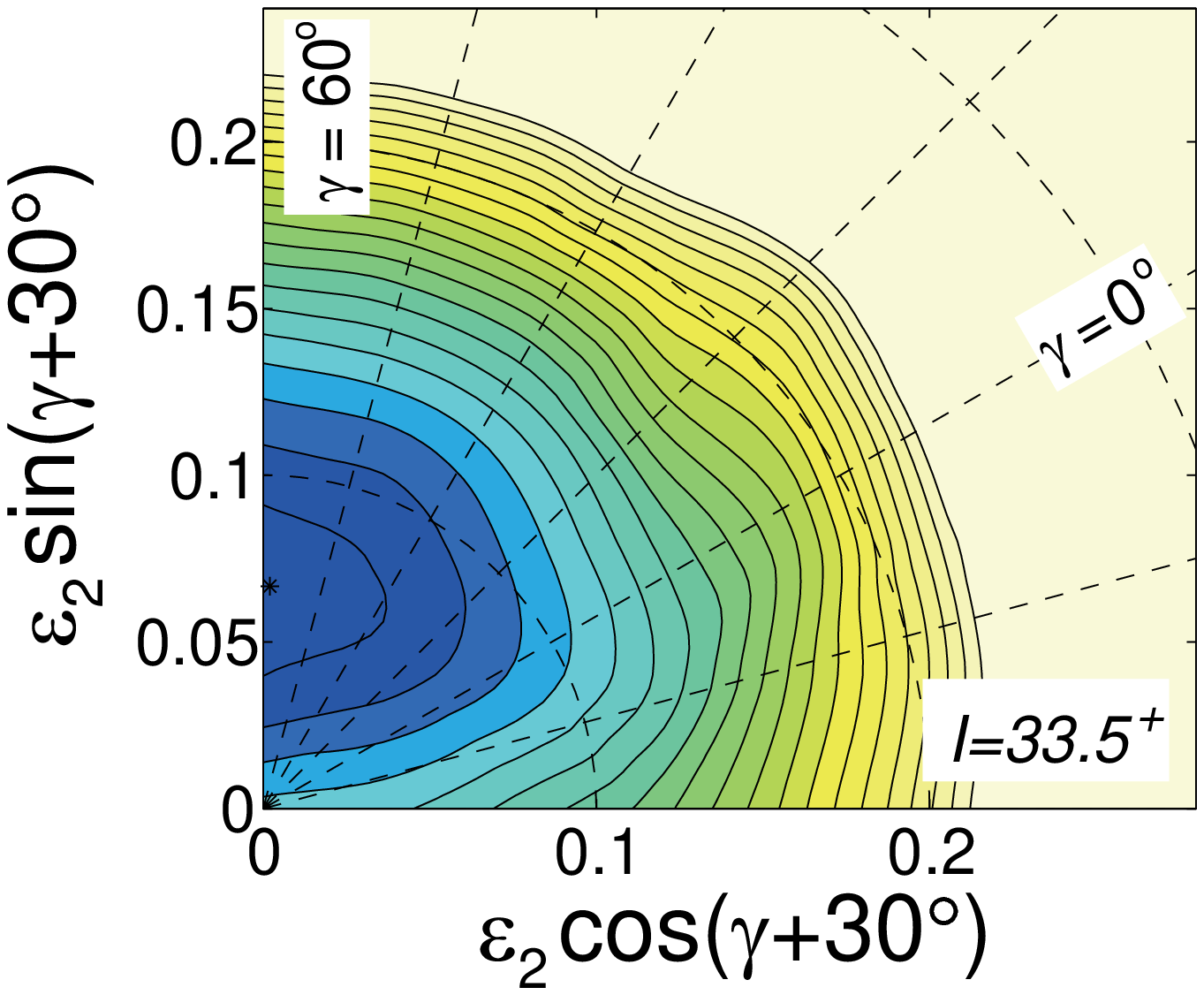}
\caption{%
(Color online) Evolution of the potential energy surfaces (PES) for the [20,2] 
configuration in $^{107}$In. The six PES figures are given in steps of $4\hbar$ 
from the near-prolate state at $I=13.5^+$ up to the terminating state at $I=33.5^+$. 
The energy difference between two neighboring equipotential lines is 0.2 MeV 
and the last equipotential line corresponds to 4.0 MeV excitation with respect 
to the minimum.}
\label{Pes}
\end{figure*}

The evolution of the potential energy surfaces for the [20,2] configuration
is shown in Fig.\ \ref{Pes}. 
This configuration displays  typical features of smooth band termination gradually 
evolving from near-prolate shape at low spin up to a non-collective oblate shape 
at the terminating state with $I^{\pi}=67/2^+$. 
These energy surfaces are very regular thus showing no disturbances, neither from 
mixing with aligned states within the same configuration (see Ref.\ \cite{Smooth-PR} 
for more details of such mechanism) nor from mixing with other collective configurations. 
This is a consequence of the fact that the nuclear system attempts to avoid a high level 
density in the vicinity of the Fermi level by a gradual adjustment of the equilibrium 
deformation with increasing spin \cite{Smooth-PR}.

Taking into account the uncertainty of spin-parity definition of the experimental 
$\Delta I=2$ band, one can consider also alternative configuration assignments. 
These are the [30,2], [20,3] and [21,2] configurations (Fig.\ \ref{In107-eld}). 
The [30,2] configuration involves an additional excitation from the proton $g_{9/2}$ 
subshell, while the [20,3] ([21,2]) configurations are built from the [20,2] 
configuration by moving the neutron from $(d_{5/2}\,g_{7/2})$ into $h_{11/2}$ 
(by moving the proton from $(d_{5/2}\,g_{7/2})$ into $h_{11/2}$). 
These changes require additional energy, and thus these configurations are less 
energetically favored than [20,2] (Figs.\ \ref{In107-eld} and \ref{exp-th}b). 
In addition, the spin of the lowest state in the band has to be changed either 
to $21/2$ or to $25/2$ if either the [30,2] or [20,3] configurations is to be 
assigned to the experimental band. 
Note that the calculated energies for the [20,3] and [30,2] configurations are 
similar (Fig.\ \ref{exp-th}b). 
Thus, the [20,3] configuration is not shown in Fig.\ \ref{exp-th}c. 
The systematics of the configuration assignments for smoothly terminating bands 
in the $N=58$ isotones (Fig.\ 20 and 23 in Ref.\ \cite{Smooth-PR}) also disfavors 
the assignment of these configurations. 
A '[21' proton configuration corresponds to $\pi(g_{9/2})^{-2}(h_{11/2})^1$, 
i.e. the first $h_{11/2}$ orbital is filled before any positive parity orbital 
above $Z=50$ is filled, which is clearly unexpected considering that the $h_{11/2}$ 
shell is located substantially higher in energy than the $g_{7/2}$ (and $d_{5/2}$) 
subshell. 
Indeed, the configurations of this type become yrast in neighboring $^{108}$Sn 
nucleus only above $I\approx 28\hbar$. Furthermore configurations of the type 
[30,2] with 3  $g_{9/2}$ proton holes have not been assigned to any smooth 
terminating band in this region while configurations with three $g_{9/2}$ neutrons 
are only energetically favoured at higher spin values, $I \approx 30-40$ 
(Ref.\ \cite{Smooth-PR}).

   One can see that it is only the [20,2] configuration which gives a reasonable 
description of the experimental data. 
Note that the energy difference between calculations and experiment for the 
[20,2] configuration (Fig.\ \ref{exp-th}c) is similar to that for other terminating 
bands in this region, see Fig. 1 of Ref. \cite{CR.06}, with a constant and 
somewhat negative value around 0.5-0.8 MeV in an extended spin range before 
termination. 
All the other curves show very different features. 
Thus, the CNS comparison with experiment clearly selects the [20,2] configuration 
as the only reasonable assignment for the observed band.\\

\section{Summary}

 In summary, a rotational band structure with ten E2 cascade transitions
has been observed in $^{107}$In. The J$^{(1)}$ and  J$^{(2)}$ moments of the 
band exhibit a gradual decrease with increasing rotational frequency, a typical 
characteristic of the smoothly terminating bands in this mass region. The 
experimental results were compared with Total Routhian Surface (TRS) and Cranked 
Nilsson-Strutinsky (CNS) calculations. In the former case, pairing was taken 
into account and in the latter, it was neglected. In both calculations, the 
configuration with positive parity and negative signature $(+, -1/2)$ was 
energetically most favored in the spin range $I = 20 - 31 \hbar$. 
According to the CNS calculations, the observed band has a [20,2] structure: 
under this configuration assignment it is one transition short of termination. 
However, the TRS calculations give some preference for a negative parity
assignment instead, corresponding to an approximate [21,2] configuration
at high spin using the CNS labels.
Further experimental investigations of this nucleus will be useful to 
observe higher spin levels and definitely fix the spin of the observed band.

%
%

\begin{acknowledgments}
The authors would like to thank Geirr Sletten and Jette S$\o$rensen 
at NBI, Denmark for preparing the targets. 
We thank the UK/France (STFC/IN2P3) Loan Pool and {\sc gammapool} European
Spectroscopy Resource for the loan of the detectors for {\sc jurogam}.
This work was supported by the Swedish Research Council, the Academy of 
Finland under the Finnish Center of Excellence Programme 
2000--2005 (Project No. 44875, Nuclear and Condensed Matter Physics 
Programme at JYFL), 
UK STFC,
the G\"oran Gustafsson foundation, the JSPS Core-to-Core Program,
International Research Network for Exotic Femto System, the European 
Union Fifth Framework Programme 
``Improving Human Potential--Access to Research Infrastructure''
(Contract No. HPRI-CT-1999-00044), 
by the U.S. Department of Energy under Grant  DE-FG02-07ER41459, 
and was also supported by a travel grant to JUSTIPEN (Japan-US Theory 
Institute for Physics with Exotic Nuclei) under U.S. Department of 
Energy Grant DE-FG02-06ER41407.
\end{acknowledgments}


\end{document}